\documentclass[
reprint,
superscriptaddress,
amsmath,amssymb,
prb
]{revtex4-2}
\usepackage{graphicx}
\usepackage{bm}
\usepackage{physics}
\usepackage{placeins}
\usepackage{braket}
\usepackage{dsfont} 
\usepackage{xcolor}
\usepackage{lipsum}
\usepackage{inputenc}
\usepackage[colorlinks=true]{hyperref}

\usepackage{ulem}  
\usepackage{cancel}
\usepackage{color}
\definecolor{d_green}{RGB}{0,128,0}

\makeatletter
\renewcommand\@make@capt@title[2]{%
  \@ifx@empty\float@link{\@firstofone}{\expandafter\href\expandafter{\float@link}}%
   {\textbf{#1}}\@caption@fignum@sep#2\quad
}%
\makeatother

\begin{document}
\title{Apparent nonlinear damping triggered by quantum fluctuations}
\author{Mario F. Gely}
\affiliation{%
Kavli Institute of NanoScience, Delft University of Technology,\\
PO Box 5046, 2600 GA, Delft, The Netherlands.
}
\author{Adri\'an Sanz Mora}
\affiliation{%
Kavli Institute of NanoScience, Delft University of Technology,\\
PO Box 5046, 2600 GA, Delft, The Netherlands.
}
\author{Shun Yanai}
\affiliation{%
Kavli Institute of NanoScience, Delft University of Technology,\\
PO Box 5046, 2600 GA, Delft, The Netherlands.
}
\affiliation{Current address: 
Institute for Quantum Computing, University of Waterloo, 200 University Avenue West, Waterloo, Ontario N2L 3G1, Canada}
\affiliation{Current address: 
Department of Physics and Astronomy, University of Waterloo, 200 University Avenue West, Waterloo, Ontario N2L 3G1, Canada
}
\author{Rik van der Spek}
\affiliation{%
Kavli Institute of NanoScience, Delft University of Technology,\\
PO Box 5046, 2600 GA, Delft, The Netherlands.
}
\author{Daniel Bothner}
\affiliation{%
Kavli Institute of NanoScience, Delft University of Technology,\\
PO Box 5046, 2600 GA, Delft, The Netherlands.
}
\affiliation{Physikalisches Institut, Center for Quantum Science (CQ) and LISA$^+$, University of T\"ubingen, Auf der Morgenstelle 14, 72076 T\"ubingen, Germany}

\author{Gary A. Steele}
\affiliation{%
Kavli Institute of NanoScience, Delft University of Technology,\\
PO Box 5046, 2600 GA, Delft, The Netherlands.
}
\date{\today}

\maketitle
\textbf{  
Nonlinear damping, the change in damping rate with the amplitude of oscillations plays an important role in many electrical~\cite{van1926lxxxviii}, mechanical~\cite{taylan2000effect} and even biological~\cite{fitzhugh1961impulses} oscillators.
In novel technologies such as carbon nanotubes, graphene membranes\cite{eichler2011nonlinear} or superconducting resonators~\cite{castellanos2007widely}, the origin of nonlinear damping is sometimes unclear.
This presents a problem, as the damping rate is a key figure of merit in the application of these systems to extremely precise sensors~\cite{chaste2012nanomechanical,hanay2012single} or quantum computers~\cite{arute2019quantum}.
Through measurements of a superconducting resonator, we show that from the interplay of quantum fluctuations and the nonlinearity of a Josephson junction emerges a power-dependence in the resonator response which closely resembles nonlinear damping.
The phenomenon can be understood and visualized through the flow of quasi-probability in phase space where it reveals itself as dephasing. 
Crucially, the effect is not restricted to superconducting circuits: we expect that quantum fluctuations or other sources of noise give rise to apparent nonlinear damping in systems with a similar conservative nonlinearity, such as nano-mechanical oscillators or even macroscopic systems.
}


Amplitude-dependent (nonlinear) damping is ubiquitous in nature.
It was famously described mathematically by van der Pol~\cite{van1926lxxxviii} in the context of his work on vacuum tube circuits~\cite{cartwbight1960balthazar}.
Now, it is used to describe the physics of a diverse set of systems, such as the rolling of ships in waves~\cite{taylan2000effect} or the nervous system~\cite{fitzhugh1961impulses}.
It has attracted recent interest due to its appearance in novel experimental platforms such as nanoscale ferromagnets~\cite{barsukov2019giant}, superconducting circuits~\cite{gao2007noise,castellanos2007widely,o2008microwave,gao2008experimental} and  nanoelectromechanical systems (NEMS)~\cite{PhysRevE.74.046619,zaitsev2012nonlinear,imboden2013observation,imboden2014dissipation} made for example from carbon nanotubes, graphene~\cite{eichler2011nonlinear,singh2016negative} or superconducting metal~\cite{yanai2017mechanical}.
In some of these systems the nonlinearity is well explained~\cite{atalaya2016nonlinear,shoshani2017anomalous,guttinger2017energy,dong2018strong}.
Most notably the saturation of two-level systems in the environment can cause negative nonlinear damping: the damping rate decreases as the power injected into the system increases~\cite{yanai2017mechanical,gao2007noise,o2008microwave,gao2008experimental}.
But the origin of an increase in damping with power in certain NEMS~\cite{eichler2011nonlinear,zaitsev2012nonlinear,imboden2013observation,imboden2014dissipation} or superconducting resonators~\cite{castellanos2007widely} remains speculative.
Understanding the origin of nonlinear damping in some of these systems is critical due to the importance of their energy damping rates in applications such as NEMS based mass sensing~\cite{chaste2012nanomechanical} or spectrometry~\cite{hanay2012single}, as well as quantum-limited amplification~\cite{vijay2011observation,aumentado2020superconducting} in superconducting quantum computers~\cite{arute2019quantum}.

We study a dephasing effect in superconducting circuits~\cite{vool2017introduction}, which phenomenologically appears as nonlinear damping when measuring the resonant response of a resonator.
Central to the observed physics is the nonlinearity induced by a Josephson junction: that the resonance frequency varies with the oscillation amplitude, which can be further approximated as a Duffing or Kerr nonlinearity~\cite{blais2020circuit}.
For this reason, the phenomena discussed here are applicable to all systems featuring a similar nonlinearity in their resonance frequency, for example the carbon-nanotubes mentioned above, or even a macroscopic mechanical pendulum.
We focus on the regime where this nonlinearity is small, as in Josephson parametric amplifiers~\cite{aumentado2020superconducting}, rather than the single-photon nonlinear regime used to construct artificial atoms in circuit quantum electrodynamics~\cite{blais2020circuit}.
Because of their small nonlinearity, such systems are often thought to be completely described by the classical Kerr oscillator~\cite{castellanos2007widely,beltran2010development}. 

Here however, we report on an effect that is not expected from the classical Kerr oscillator.
More specifically, we present a phenomenon triggered by the interplay between the quantum noise and Kerr anharmonicity of the oscillator, which closely resembles nonlinear damping in the steady-state response of the oscillator.
The apparent nonlinear damping is first experimentally characterized by probing the frequency response of the resonant circuit.
Our observations are then accurately described by a quantum theory of a damped driven Kerr oscillator devoid of ad-hoc nonlinear damping, but which takes into account the effect of quantum noise.
Moreover, focusing on an oscillator steady-state below its bistability threshold, a Gaussian state approximation~\cite{degenfeld2015self} allows us to demonstrate that, in a close vicinity of the resonance, the expected amplitude of oscillations is akin to that of a driven classical Kerr oscillator with nonlinear damping.
Finally, we provide an intuitive picture in which the phenomenon can be understood as the oscillator experiencing dephasing induced by its own photon shot noise.

\begin{figure}[t!]
\includegraphics[width=0.48\textwidth]{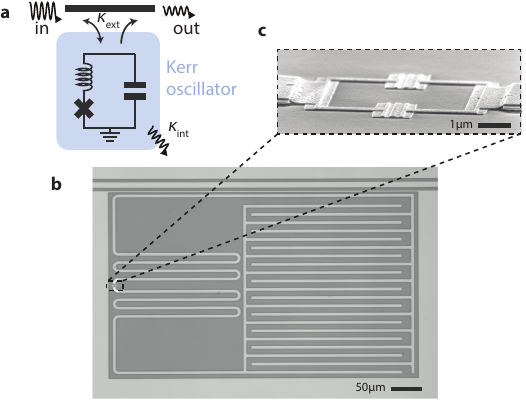}
\caption{
	\textbf{Superconducting Kerr oscillator circuit.}
	\textbf{a},
	The Kerr oscillator is constructed from an inductor, a capacitor and a SQUID (which behaves as and is depicted by a single Josephson junction), and is side-coupled to a transmission line with a coupling rate $\kappa_\text{ext}$.
	The circuit undergoes internal damping at a rate $\kappa_\text{int}$.
	\textbf{b},
	Optical micrograph of the device, where light gray corresponds to superconducting molybdenum-rhenium, and dark gray to the insulating silicon substrate.
	An interdigitated capacitor on the right is connected to a meandering inductor on the left.
	The circuit couples to a transmission line (coplanar waveguide) at the top.
	\textbf{c},
	Scanning electron micrograph of the SQUID: two aluminum/aluminum-oxide Josephson junctions connected in parallel. 
	As the flux threading the SQUID is fixed, it effectively behaves in this context as a single junction.
}
\label{fig:1}
\end{figure}

\begin{figure}[t!]
\includegraphics[width=0.48\textwidth]{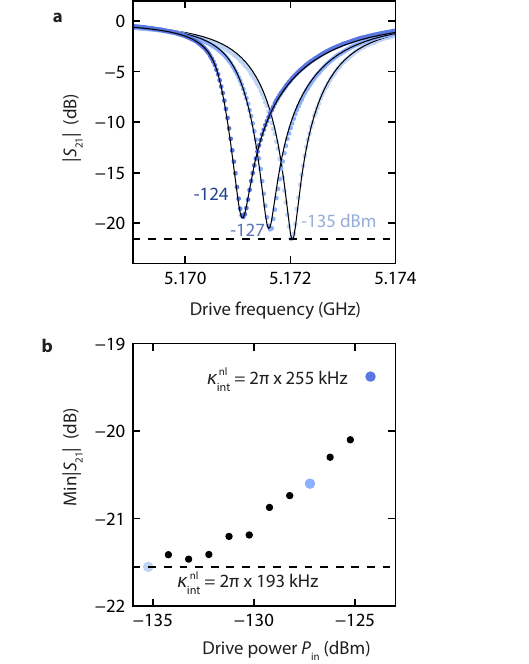}
\caption{
	\textbf{Observation of a resonator steady-state response suggesting nonlinear damping.}
	\textbf{a},
	Measured transmission magnitude $|S_\text{21}|$ (dots) for different drive powers.
	Whilst the shift in resonance frequency is expected from a classical analysis of the damped driven Kerr oscillator using Eq.~(\ref{eq:main_classical_steady_state_equation}), Min$|S_\text{21}|$ is expected to remain constant (dashed line).
	\textbf{b},
	Measured Min$|S_\text{21}|$ (dots) as a function of drive power.
	Eq.~(\ref{eq:main_classical_steady_state_equation}) yields Min$|S_\text{21}| = |1-\kappa_\text{ext}/(\kappa_\text{int}+\kappa_\text{ext})|$, suggesting a damping rate which increases with power $\kappa_\text{int}\rightarrow\kappa_\text{int}^\text{nl}(|a|)$ from $\kappa_\text{int}^\text{nl}=2\pi\times 193$ kHz to $\kappa_\text{int}^\text{nl}=2\pi\times 255$ kHz.
	Indeed, adding nonlinear damping $\kappa_\text{int}^\text{nl}(|a|)$ to Eq.~(\ref{eq:main_classical_steady_state_equation}) leads to theoretical predictions (solid lines in \textbf{a}) in good agreement with the data.
	At the three highlighted points, the expectation values of photon number $|a|^2$ (where the minimum of $|S_\text{21}|$ is achieved) are 1.1, 6.8 and 13.2.
}
\label{fig:2}
\end{figure}
	%
	The circuit used in this experiment (Fig.~\ref{fig:1}) is constructed from an inductor, capacitor and superconducting quantum interference device, or SQUID.
	The SQUID is flux-biased to its sweet spot (integer flux quantum), and behaves as a single Josephson junction~\cite{koch2007charge}.
	The junction induces an anharmonicity of strength $K = 2\pi\times 80$ kHz five orders of magnitude smaller than the resonance frequency $\omega_\text{r} = 2\pi\times 5.17$ GHz.
	%
	The cosine potential of the junction is accurately described in this limit $K\ll\omega_\text{r}$ by the Kerr effect in the Hamiltonian~\cite{gely2020qucat}
	\begin{equation}
	 	\hat H = \hbar\bigg(\omega_\text{r}\underbrace{-\tfrac{K}{2}\hat a^\dagger\hat a}_\text{Kerr}-\tfrac{K}{2}\bigg)\hat a^\dagger\hat a\ ,
	 	\label{eq:main_Hamiltonian}
	\end{equation}
	where $\hat a$ is the annihilation operator for photons in the circuit.
	Intuitively, the junction is acting as an inductor, with an inductance which increases with the number of photons $\hat a^\dagger\hat a$ in the circuit.
	As a consequence, the resonance frequency of the circuit is lowered with each added photon, labeled as the Kerr term in Eq.~(\ref{eq:main_Hamiltonian}).

	The circuit undergoes internal damping, losing energy at a rate $\kappa_\text{int} = 2\pi\times 186$ kHz.
	This is typically due to losses in the different dielectric materials traversed by the electric fields~\cite{wang2015surface}.
	Additionally, the circuit is coupled to a transmission line, through which we drive the circuit with a microwave signal.
	Conversely, the transmission line leads to energy leaking out of the circuit, which is characterized by an external damping rate ${\kappa_\text{ext} =2\pi\times  2.1 \, {\rm MHz}}$.
	As a consequence, the total damping rate and spectral linewidth $\kappa = \kappa_\text{int}+\kappa_\text{ext}$ is much larger than the shift in resonance frequency $K$ due to an added photon: $\kappa\gg K$.
	The circuit is thus far from the regime of superconducting qubits~\cite{blais2020circuit}.
	We will call it a Kerr oscillator and first attempt to describe its behavior following the classical equation for the steady-state amplitude of its oscillations $a$ 
	\begin{equation}
	\left(i\Delta-iK |a|^2+\frac{\kappa}{2}\right)a  =\epsilon\ .
	\label{eq:main_classical_steady_state_equation}
	\end{equation}
	Here $\Delta = \omega_\text{r}-\omega_\text{d}$ is the detuning of the driving frequency $\omega_\text{d}$ to the resonance frequency $\omega_\text{r}$, and the strength of the drive $\epsilon = \sqrt{\kappa_\text{ext}P_\text{in}/(2\hbar\omega_\text{r})}$ is given by $P_\text{in}$ the power of the drive impinging on the device.


	The circuit is made by patterning a thin film of sputtered molybdenum-rhenium alloy on silicon, and subsequently fabricating the aluminum/aluminum-oxide tunnel junctions 
	(see Supplementary Information Sec. S7).
	The device is thermally anchored to the $\sim$ 20 milliKelvin stage of a dilution refrigerator, and the input (output) microwave wiring is attenuated (isolated) to lower its microwave mode temperature, such that the average number of photons $n_\text{th}$ excited by thermal energy is negligible.
	The transmission coefficient $S_\text{21} =  1-\kappa_\text{ext}a/(2\epsilon)$ is then measured using a vector network analyzer (VNA) for varying microwave power (see Fig.~\ref{fig:2}\textbf{a}).



	We note an increase in both the detuning $\Delta_\text{min}$ which minimizes transmission, and the value of the minimum Min$|S_\text{21}|$.
	The classical prediction $\Delta_\text{min} = K\alpha^2$ resulting from Eq.~(\ref{eq:main_classical_steady_state_equation}) -- where  $\alpha = 2\epsilon/\kappa$ is the expected maximum amplitude -- accurately matches the shift of the resonance.
	%
	However, by plugging the maximum amplitude $\alpha$ into the expression for $S_\text{21}$, we obtain a constant value for Min$|S_\text{21}| =  |1-\kappa_\text{ext}/\kappa|$ (dashed line in Fig.~\ref{fig:2}\textbf{a}), which disagrees with the measurement.

	In a classical approach to the problem, a power-dependence of the internal damping rate therefore has to create this change.
	Since $\kappa_\text{ext}$ is determined by the geometry of the circuit, it should remain unchanged by the power of the drive.
	For $\kappa_\text{ext}/\kappa$ to vary and produce the observed change in Min$|S_\text{21}|=|1-\kappa_\text{ext}/(\kappa_\text{ext}+\kappa_\text{int})|$, the internal damping should increase as the drive power increases.
	At the highest drive power for which data is displayed (\({P_{\rm in}=-124 \, {\rm dBm}}\)), the internal damping rises to ${2\pi\times 255 \, {\rm kHz}}$.
  We note that this power lies below the bistability threshold (see Supplementary Information Secs.~S3~and~S4F).
	Such nonlinear damping can be included in the model of Eq.~(\ref{eq:main_classical_steady_state_equation}) through $\kappa_\text{int}\rightarrow\kappa_\text{int}^\text{nl}=\kappa_\text{int}+\gamma|a|^2$.
	We fit a solution of the resulting equation to the data (see Supplementary Information Sec. S2), observing good agreement (Fig.~\ref{fig:2}) for $\gamma=2\pi\times 5.02$ kHz.
	%
	%

	%
	Whilst providing an accurate model for our observations, adding ad-hoc nonlinear damping offers no explanation as to the physical mechanism underlying the effect.
	Usually, the most prominent source of nonlinear damping in superconducting circuits is the saturation of two-level systems (TLSs) in the environment~\cite{gao2007noise,o2008microwave,gao2008experimental}.
	However, with increasing driving power, the saturation of TLSs will result in a \textit{decrease} of the internal damping rate, whilst we observe the opposite.
	Here, we show that in our system this nonlinear damping behavior can be explained purely by dephasing triggered by the joint action of the intrinsic quantum noise and the Kerr anharmonicity of the oscillator.

\begin{figure}[t!]
\includegraphics[width=0.48\textwidth]{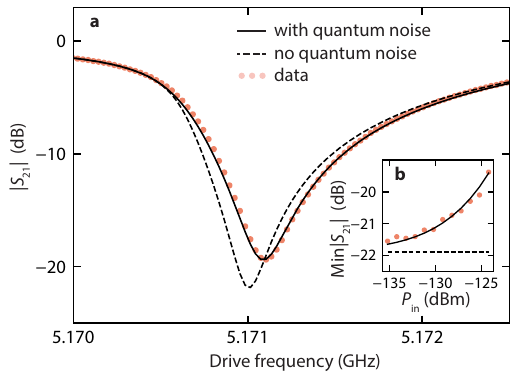}
\caption{
	\textbf{Apparent nonlinear damping triggered by quantum noise.}
	The experimental results (dots) are compared here to a model with and without quantum noise (full and dashed line respectively).
	\textbf{a}, Example of experimental and theoretical $|S_\text{21}|$ at the input power $P_\text{in} = -124$ dBm.
	\textbf{b}, As the power varies the model without quantum noise fails to capture the power-dependent depth of the response, which is accurately reproduced when quantum noise is introduced.
	}
	\label{fig:main_quantum_fit}
	\end{figure}

\begin{figure*}[t!]
\includegraphics[width=0.96\textwidth]{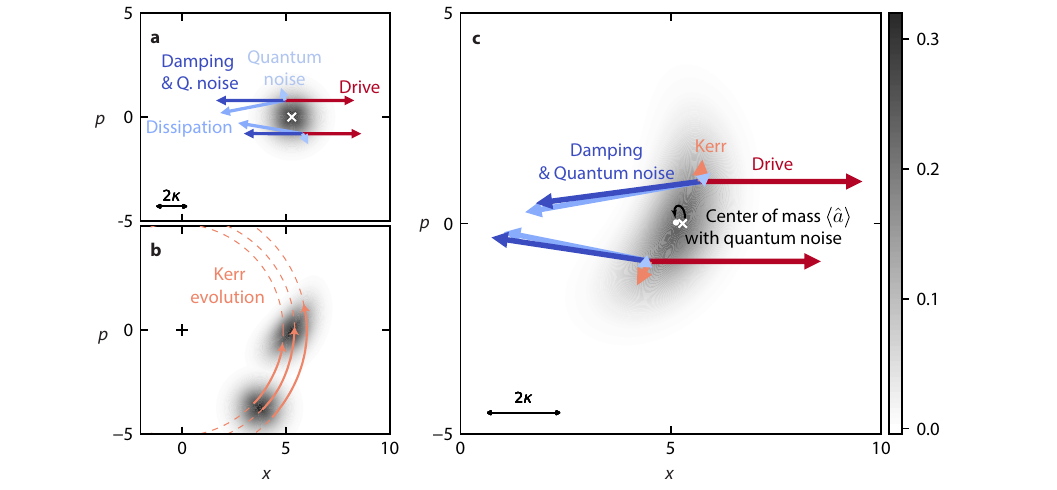}
\caption{
	\textbf{Phase space picture of apparent damping.}
	Here phase space operators are defined by $\hat a=\left(\hat x +i\hat p\right)/\sqrt{2}$, see Supplementary Information Sec. S5A for further details.
	\textbf{a},
	Wigner distribution of the steady-state in absence of Kerr nonlinearity (driven at $\omega_\text{r}$ with $P_\text{in}=-124$ dBm).
	The balance between quantum noise, damping and drive is shown by vectors corresponding to Wigner currents.
	\textbf{b}, 
	Growth of phase uncertainty of a coherent state under Kerr nonlinearity.
	The \textit{amplitude}-dependent \textit{resonance frequency} (Kerr effect) translates to a \textit{radius}-dependent \textit{rotation} around the origin.
	The center of mass of the distribution rotates at a frequency $K\alpha^2$.
	In a frame rotating at that frequency, the effect of the Kerr nonlinearity is to increase the uncertainty in phase (this is the frame adopted in panel \textbf{c}).
	The larger the uncertainty in phase (the extreme case being a ring around the origin), the closer the center of mass of the distribution gets to the origin (i.e. $|\langle\hat a\rangle|\rightarrow 0$).
	This is the first contribution (\textit{effect A}) to a reduced resonant amplitude's magnitude $|\langle\hat a\rangle|$.
	\textbf{c}, 
	Wigner distribution of the steady-state with Kerr nonlinearity (at minimum $|S_\text{21}|$ with $P_\text{in}=-124$ dBm).
	The Kerr effect is eventually balanced by the damping, quantum noise and drive.
	Since the drive now opposes both damping and Kerr effect, it is less effective at opposing the damping and driving the state away from the origin (compared to panel \textbf{a}).
	This brings the distribution closer to the origin, and constitutes the second contribution (\textit{effect B}) to a reduced resonant amplitude's magnitude $|\langle\hat a\rangle|$.
	The center of mass ($\langle \hat a\rangle$) (white dot) is compared to the classical steady-state (white cross).
	Since the Wigner current of the Kerr effect grows with the amplitude squared $|\langle\hat a\rangle|^2\propto\epsilon^2$ and the drive and dissipation currents grow with $\epsilon$ and $|\langle\hat a\rangle|$ respectively, the reduction in $|\langle\hat a\rangle|$ does not linearly follow the driving strength $\epsilon$ 
	(see Supplementary Information Sec. S5B).
	}
\label{fig:main_wigner}
\end{figure*}

	%
	We first show that approaching the problem quantum mechanically, without adding nonlinear damping, perfectly describes our measurements.
	The effect of quantum noise is included in the model through the steady-state Lindblad equation
	\begin{equation}
		i\big[\hat H/\hbar-\omega_\text{d}\hat a^\dagger\hat a+i\epsilon(\hat a^\dagger-\hat a),\hat \rho\big]=\kappa\big(2\hat a\hat \rho\hat a^\dagger-\hat \rho\hat a^\dagger\hat a-\hat a^\dagger\hat a\hat \rho\big)/2\ ,
		\label{eq:main_lindblad}
	\end{equation}
	where $\hat \rho$ is the density matrix describing the steady-state of the oscillator.
	By numerically solving this equation for varying drive strengths and frequencies, the resulting amplitude $\langle\hat a\rangle = \text{Tr}(\hat a\hat\rho)$ is used to obtain $S_\text{21}$.
	With only the circuit parameters as free variables, and notably a constant value for the internal damping, this model is fitted to all $S_\text{21}$ traces 
	(see Supplementary Information Sec. S2), revealing excellent agreement to the data (Fig.~\ref{fig:main_quantum_fit}).
	Note that we recently became aware of an analytical solution to this Lindblad equation~\cite{drummond1980quantum,kheruntsyan1999wigner}, which may have simplified our approach.
	In Fig.~\ref{fig:main_quantum_fit}, we compare this quantum model to the classical model: the solution to Eq.~(\ref{eq:main_classical_steady_state_equation}), which features neither nonlinear damping nor quantum noise.
	The only difference between the quantum model -- which predicts the increase in Min$|S_\text{21}|$ -- and that of Eq.~(\ref{eq:main_classical_steady_state_equation}) -- which predicts a constant Min$|S_\text{21}|$ -- lies in the value of the commutator $[\hat a,\hat a^\dagger]$.
	In fact, by taking the trace Tr$(\hat a\ \cdot\ )$ of Eq.~(\ref{eq:main_lindblad}), and assuming the amplitude to be a complex number $\hat a\rightarrow a$ such that $[a, a^*]=0$, we arrive at Eq.~(\ref{eq:main_classical_steady_state_equation}).
	Quantum noise can therefore lead to the entirety of the change in Min$|S_\text{21}|$.
	Thermal noise could lead to a similar effect, but is expected to be negligible in our experiment (see Supplementary Information Sec. S7).

	Beyond describing the data, this model can lead to a nonlinear damping equation for the expectation value $\langle \hat a\rangle$.
	In the
	Supplementary Information (Sec. S4),
	we derive an  analytical formula that captures the behavior of the steady-state response that is in good agreement with the numerical simulations.
	We find that in a resonance scenario, whenever the Kerr effect and thermal noise have only a perturbative effect on the system, the corresponding steady-state expectation value for the amplitude $\langle\hat a\rangle$ matches that of a classical non-linearly damped driven classical Kerr oscillator.
	That is, the steady-state amplitude is ruled by an equation resembling Eq.~(\ref{eq:main_classical_steady_state_equation}), but where the quantum and thermal noise lead to a nonlinear damping coefficient $\kappa+\gamma|\langle\hat a\rangle|^2$ with
	\begin{equation}
		\gamma = \frac{4K^2}{\kappa}\left(n_\text{th}+\frac{1}{2}\right)\ .
	\end{equation}
	Here the familiar $+\frac12$ stems from quantum noise, which has the same effect as half a quantum of thermal noise.
	The fact that the nonlinear damping model and the quantum model are both able to describe our measurements is therefore not coincidental: whilst there is no microscopic process leading to nonlinear damping (i.e. loss of energy), there is apparent nonlinear damping in the equation for $\langle\hat a\rangle$ when accounting for the presence of quantum noise.
	Similar results were derived for the classical~\cite{dykman1971classical} and quantum~\cite{dykman1973quantum} spectrum of undriven oscillators, and also in work studying the spectrum of a probe field in the presence of a strong pump field~\cite{dykman2012periodically}. 
	The difference here is that we are instead interested in the power-dependence of the scattering parameter $S_\text{21}$.


	We now provide an intuitive explanation as to why there is a decrease in the amplitude of oscillations $\langle\hat a\rangle$ -- leading to an increase in the minimum of $|\langle\hat S_\text{21}\rangle| =  |1-\kappa_\text{ext}\langle\hat a\rangle/(2\epsilon)|$ -- when quantum noise is considered.
	Because of the Kerr nonlinearity, the uncertainty in the photon number operator $\hat a^\dagger \hat a$, translates to uncertainty in the resonance frequency of the oscillator $\omega_\text{r}-K\hat a^\dagger \hat a/2$ (see Hamiltonian of Eq.~(\ref{eq:main_Hamiltonian})).
	This has two consequences.
	\textit{Effect A}: the signal leaking out of the oscillator into the transmission line inherits the frequency fluctuations of the oscillator.
	Since we are measuring a single frequency component with our VNA, we will measure a signal of smaller amplitude.
	\textit{Effect B}: the driving is less effective at exciting the oscillator because the resonance frequency of the oscillator is fluctuating and no driving frequency will lead to resonant driving.
	The average number of photons in the oscillator will decrease.
	%

	This interpretation can be more thoroughly explored in phase space, by making use of the Wigner distribution and Wigner current~\cite{isar1996phase,stobinska2008wigner,oliva2019quantum,friedman2017wigner}.
	%
	We introduce $\hat x =(\hat a+\hat a^\dagger)/\sqrt{2}$ and $\hat p = -i(\hat a-\hat a^\dagger)/\sqrt{2}$, such that the amplitude $\langle\hat a\rangle$ is given by the center of mass of the distribution through $\langle\hat a\rangle=\sqrt{2}\iint \text{d}x\text{d}p(x+ip)W$.
	The Wigner current $\vec J$, governs the dynamics of the Wigner function $W$ through the continuity equation $\partial_t W + \vec \nabla \vec J = 0$.
	It provides an intuitive visualization of the flow of quasi-probability in phase space.

	As a pedagogical starting point, we show in Fig.~\ref{fig:main_wigner}\textbf{a} the distribution and different contributions to the current for a coherent state of amplitude $\alpha$.
	This state corresponds to the steady-state that would be reached in our resonantly driven system without Kerr nonlinearity.
	The damping tends to bring each point of the distribution back to the origin.
	The drive however, is sensitive to phase and acts in a single direction.
	These two currents are balanced by the quantum noise, which creates a diffusion of the quasi-probability. 

	In Fig.~\ref{fig:main_wigner}\textbf{b}, we look at how the Kerr effect deforms the same coherent state, with the damping, driving, and noise temporarily inactive.
	We see the consequence of the amplitude-dependent resonance frequency of a damped driven quantum Kerr oscillator.
	In phase space, the resonance frequency sets the rate at which a point rotates around the origin.
	And the amplitude is given by the distance to the origin.
	The resulting deformation of the coherent state does not bring any point in phase space closer to the origin (total energy, or photon number, remains constant).
	The center of mass, however, will move closer to the origin.
	To be convinced of the latter, one can imagine the extreme case of the Kerr effect deforming the coherent state into a ring circling the origin, so that center of mass would be the origin, and $|\langle\hat a\rangle| =0$.
	This mechanism for reducing $|\langle \hat a\rangle|$ corresponds to \textit{effect A} previously discussed.

	In the steady-state of our experiment, simulated in Fig.~\ref{fig:main_wigner}\textbf{c}, the evolution of the Kerr effect is eventually balanced by the other currents.
	Due to the large spread of the state in phase, the diffusion induced by quantum noise is weaker, and the damping current further misaligned with the drive compared to Fig.~\ref{fig:main_wigner}\textbf{a}.
	Since the drive is not parallel to the combined currents of damping and noise, it is less effective at countering them, so less effective at driving the system.
	Or in other words, in addition to countering the damping, the drive also has to counter the evolution of the Kerr effect.
	As a consequence, the average photon-number tends to decrease, which is the second contribution to a lower amplitude \textit{(effect B)}.
	In the Supplementary Information (Sec. S5B), we elaborate on why this decrease in amplitude is nonlinear with driving power.

	Using a Gaussian state approximation
	(see Supplementary Information Sec. S4)
	, we are able to weigh the influence of \textit{effect A} and \textit{effect B} in reducing the value of the resonant amplitude $\langle\hat a\rangle$.
	We rely on an analytical comparison of the corresponding amplitude's magnitude $|\langle\hat a\rangle|$ and photon number $\langle\hat a^\dagger\hat a\rangle$, and the fact that \textit{effect A} does not affect the photon number, whereas \textit{effect B} reduces the amplitude by reducing the photon number from its expected value for a coherent state $\sqrt{\langle\hat a^\dagger\hat a\rangle} = |\langle\hat a\rangle|$.
	With respect to a coherent state of amplitude $\alpha$, the reduction in $\sqrt{\langle\hat a^\dagger\hat a\rangle}$ corresponds to half the reduction in $|\langle\hat a\rangle|$ in our system (without thermal noise).
	This means that a reduction in photon number is responsible for only half of the observed effect, indicating that half of the increase in ${\rm Min}|S_\text{21}|$ can be attributed to \textit{effect A}, and half to \textit{effect B}.
	The same conclusions can be drawn with thermal noise (assuming $n_\text{th}\ll\alpha^2$).

	Whilst we have focused on the case $n_\text{th}\ll\alpha^2$, where the damping seems to increase with the amplitude of oscillations, the opposite regime $n_\text{th}\gg\alpha^2$ has already been explored experimentally~\cite{eichler2013symmetry,maillet2017nonlinear} and bears some common features with this work.
	When thermal fluctuations dominate, the state of the oscillator is well described as a statistical mixture of oscillatory amplitudes, each shifting the resonance frequency by a different amount given by the Duffing nonlinearity.
	This results in a broadening of the resonance line-shape when the oscillator is probed, which has been phenomenologically interpreted as an increase in damping, for example in carbon nanotubes~\cite{eichler2013symmetry}.
	This picture even extends to the case $n_\text{th}\ll\alpha^2$ where a residual broadening persists due to quantum heating of the oscillator by the driving field~\cite{dykman2011quantum}.

	Finally, we note that for all driving strengths featured in our measurements, quadrature squeezing occurs along an axis $u = \cos(\theta)x+\sin(\theta)p$, rotated by an angle $\theta$ with respect the the $x-$axis.
	At the highest driving power (Figs.~\ref{fig:main_quantum_fit},\ref{fig:main_wigner}), the most highly squeezed quadrature is characterized by $\theta \simeq -0.11\pi$, where the uncertainty $\Delta u$ is $83$ \% of $\Delta x$ for a coherent state.


In conclusion, we have shown how the combination of Kerr nonlinearity and noise, and in particular quantum noise, leads to a dephasing that can manifest in the same way as nonlinear damping.
Crucially, our findings are not limited to the case of superconducting resonators.
Indeed, preliminary calculations based on our analytical model indicate that this effect has the correct order of magnitude to play a role in the nonlinear damping observed in NEMS systems~\cite{eichler2011nonlinear} -- however driven by thermal rather than quantum noise.
We are therefore confident that this phenomenon can play a valuable role in identifying the nature of nonlinear damping effects in a broader class of systems, such as NEMS or other Josephson circuits, which will be critical to their use in emerging technologies ranging from carbon nanotube sensors to superconducting quantum computing.

\textbf{Data and code availability}
The experimental data and software used to generate the figures in the main text and the supplementary information are available in Zenodo with the DOI identifier 10.5281/zenodo.4565179.





\textbf{Acknowledgments}
We thank M. Kounalakis for discussions.
This work was supported by the European Union’s Horizon 2020 research and innovation programme under grant agreements 681476 - QOM3D and 828826 - Quromorphic, and by the research programme of the Foundation for Fundamental Research on Matter (FOM), which was part of the Dutch Research Council (NWO).

\textbf{Author contributions}
S.Y. performed the design and fabrication of the device, the measurements, and the initial data-analysis.
D.B. and M.F.G. analyzed the data in the context of a classical nonlinear model.
M.F.G. and R.S. analyzed the data using a quantum model, including fitting and the phase space interpretation, with A.S.M. carrying out the theoretical proof that the quantum model leads to the nonlinear damping equation.
M.F.G. wrote the manuscript with contributions from all authors.
G.A.S. supervised the project.



\FloatBarrier
\clearpage
\onecolumngrid

\centerline{\huge Supplementary Information}

\makeatletter
   \renewcommand\l@section{\@dottedtocline{2}{1.5em}{1em}}
   \renewcommand\l@subsection{\@dottedtocline{2}{3.5em}{1em}}
   \renewcommand\l@subsubsection{\@dottedtocline{2}{5.5em}{1em}}
\makeatother

\renewcommand{\thesection}{\arabic{section}} 
\let\oldaddcontentsline\addcontentsline
\renewcommand{\addcontentsline}[3]{}
\let\addcontentsline\oldaddcontentsline
\renewcommand{\theequation}{S\arabic{equation}}
\renewcommand{\thefigure}{S\arabic{figure}}
\renewcommand{\thetable}{S\arabic{table}}
\renewcommand{\thesection}{S\arabic{section}}
\setcounter{figure}{0}
\setcounter{equation}{0}
\setcounter{section}{0}

\section{Theoretical description}
\label{sec:S_theoretical_description}
A Hamiltonian describing the series assembly of a inductor $L$, capacitor $C$ and junction with Josephson inductance $L_\text{J}$ can be entirely determined with only the knowledge of the admittance $Y(\omega) = 1/Z(\omega)$ across the Josephson junction, if we replace the latter by a linear inductor $L_\text{J}$.
This approach assumes weak anharmonicity and damping of the circuit, and is often referred to as black-box quantization~\cite{nigg2012black,gely2020qucat}.
The Hamiltonian writes
\begin{equation}
    \hat{H}_\text{bare}/\hbar = \left(\omega_\text{r} -K\right) a^\dagger a-\frac{K}{2} a^\dagger a^\dagger a  a\ ,
    \label{eq:hamiltonian}
\end{equation}
where $\omega_\text{r}$ satisfies $Y(\omega_\text{r})=0$
\begin{equation}
    \omega_\text{r} = \frac{1}{\sqrt{(L+L_\text{J})C}}
\end{equation}
and the Kerr constant is given by
\begin{equation}
    \hbar K = \frac{2e^2}{L_\text{J}\omega_\text{r}^2(\text{Im}Y'(\omega_\text{r}))^2}= \frac{e^2}{2C}\left(\frac{L_\text{J}}{L+L_\text{J}}\right)^3\ .
    \label{eq:anharmonicity}
\end{equation}
The weak anharmonicity assumption which leads to this Hamiltonian writes $K \ll \omega_\text{r}$.
This circuit loses energy through resistive losses at a rate $\kappa_\text{int}$, and can exchange energy with a transmission line at a rate $\kappa_\text{ext}$.
The total rate at which the circuit loses energy is then $\kappa = \kappa_\text{int}+\kappa_\text{ext}$.
On one end of the transmission line, we feed a coherent signal with power $P_\text{in}$ oscillating at $\omega_\text{d}$.
Following quantum input-output theory~\cite{vool2017introduction}, the dynamics of $\hat a(t)$, in a frame rotating at $\omega_\text{d}$, is given by
\begin{equation}
    \frac{d}{dt}\hat a(t) = -i\left(\Delta-K\hat a(t)^\dagger\hat a(t)\right)\hat a(t)-\frac{\kappa}{2}\hat a(t)+\epsilon-\sqrt{\kappa}\hat s\ ,
 \label{eq:langevin}
\end{equation}
where $\Delta = \omega_\text{r}-K-\omega_\text{d}$, and the strength of the drive is characterized by $\epsilon = \sqrt{\kappa_\text{ext}P_\text{in}/(2\hbar\omega_\text{r})}$.
Note the factor $2$ in the denominator which corresponds to the fact that there are two directions of propagation in the feedline and that only one is occupied by the driving signal.
The term $\hat s$ corresponds to both thermal noise and quantum vacuum noise.
We assume it to be well described by quantum white noise, a stationary random process which is characterized by its 0 mean $\langle \hat s\rangle=0$ and the correlation functions
\begin{equation}
    \begin{split}
        \langle\hat s (t)\hat s^\dagger(t')\rangle&=\langle\hat s^\dagger(t')\hat s (t)\rangle+\delta(t-t')=[n_\text{th}+1]\delta(t-t')\\
        \langle\hat s (t)\hat s(t')\rangle&=\langle\hat s^\dagger(t')\hat s^\dagger (t)\rangle=0\ .
    \end{split}
    \label{eq:s_correlations}
\end{equation}
Here $n_\text{th}$ corresponds to the average number of excitations induced by the thermal environment at a temperature $T$
\begin{equation}
    n_\text{th} = \frac{1}{e^{\frac{\hbar\omega_\text{r}}{k_\text{B} T}}-1}\ ,
\end{equation}
with $k_\text{B}$ corresponding to Boltzmann's constant, to which is added ``$+1$'' corresponding to the quantum vacuum fluctuations.
As described in Sec.~\ref{sec:S_fab_setup}, thermal noise is heavily attenuated such that the resonator has a thermal occupation $n_\text{th} < 0.05$, a much smaller source of noise than quantum fluctuations.
We may thus safely make the approximation $n_\text{th}\simeq 0$ when describing the experiment.
The $S_\text{21}$ parameter is obtained from $\langle \hat a(t)\rangle$ as
\begin{equation}
    S_\text{21} =  1-\frac{\kappa_\text{ext}}{2\epsilon}\langle \hat{a}\rangle\ ,
\end{equation}
the factor $2$ again reflecting that only half of the signal emitted by the circuit will travel towards the receiver of the VNA.
As an alternative to the Langevin equation, one may also formulate the problem in terms ofa Lindblad master equation~\cite{bishop2010circuit}
\begin{equation}
\begin{split}
    \frac{\partial\hat\rho}{\partial t}=&-i\left[\Delta\hat a^\dagger \hat a -\frac{K}{2}\hat a^\dagger\hat a^\dagger \hat a \hat a +i\epsilon(\hat a^\dagger-\hat a),\hat \rho\right]\\
     &+ \kappa (n_\text{th}+1) D(\hat a )\hat \rho+ \kappa n_\text{th} D(\hat a^\dagger )\hat \rho\ ,
 \end{split}
 \label{eq:lindblad}
\end{equation}
governing the density matrix of the system $\hat \rho$, where
\begin{equation}
  D(\hat L)\hat \rho = \hat L \hat \rho \hat L^\dagger - \{\hat L\hat L^\dagger,\hat \rho\}/2\ .
\end{equation}

\section{Data processing and fitting}
\label{sec:S_data_processing_fitting}

Even at lowest driving power, the response of the device does not perfectly fit to a Lorentzian curve, indicating the presence of additional resonances in the measurement chain which could not be calibrated out experimentally.
To eliminate these, as well as the change in phase length of the cabling with frequency, we subtract (divide) an affine function of frequency to the measured phase (amplitude).
\begin{figure}[htb]
  \begin{centering}
\includegraphics[width= 0.83\linewidth]{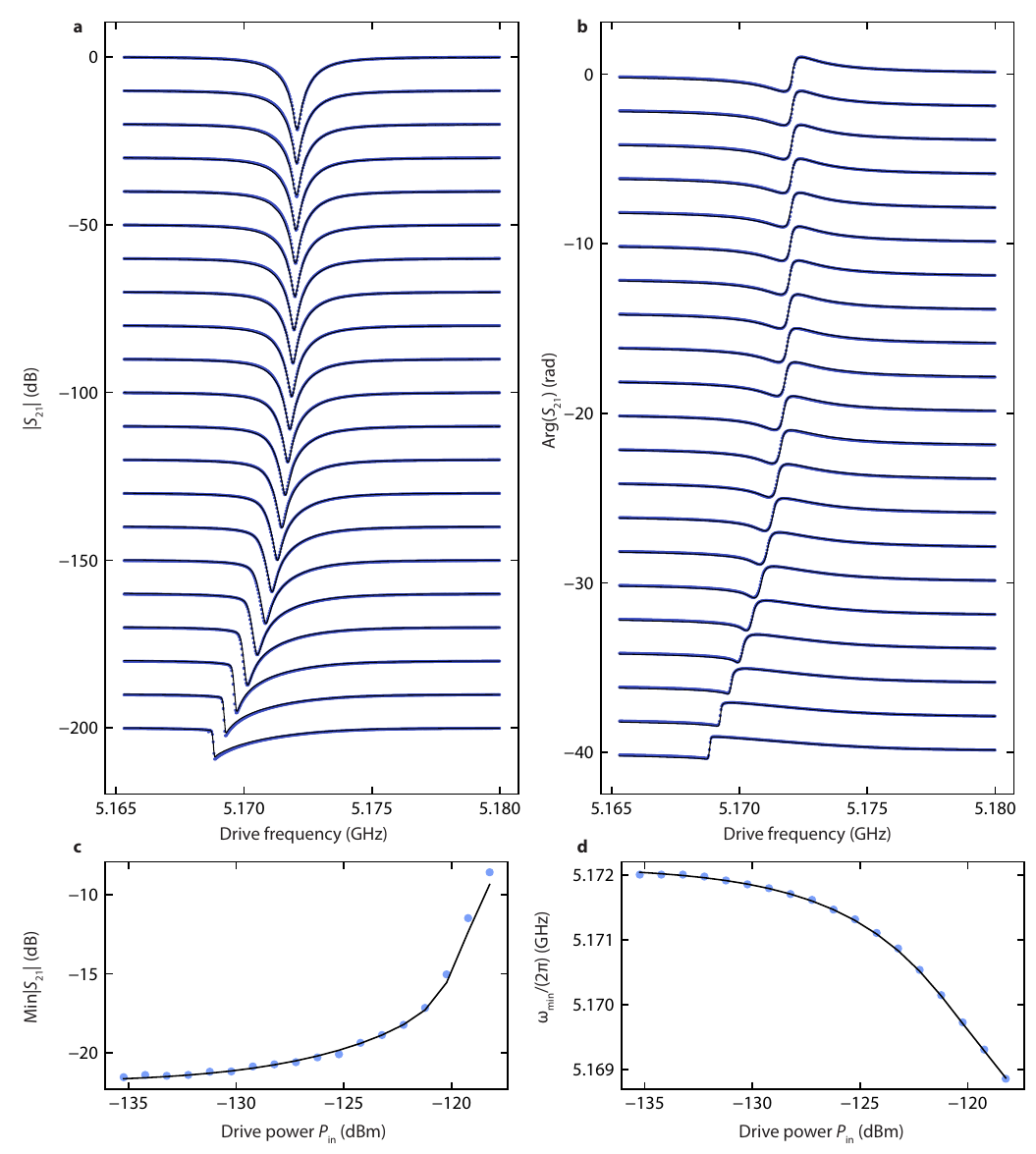} 
\par\end{centering}
\caption{\label{fig:lindblad-transmision-fits} 
  \textbf{Transmission coefficient of the superconducting resonant circuit.}
  Measurements of the transmission coefficient \(S_{\rm 21}\), processed following Sec.~\ref{sec:S_data_processing_fitting}, are used in a curve fitting routine that seeks, for each driving power in the dataset and \(n_{\rm th}=0\), a numerical solution of the steady-state equation~(\ref{eq:main_lindblad}) of the damped driven quantum Kerr oscillator with the set of oscillator parameters that fits best to the measured data in accordance to a minimization of the error in: \textbf{a,} the magnitude \(|S_{\rm 21}|\) and \textbf{b,} phase \({\rm Arg}(S_{\rm 21})\) of \(S_{\rm 21}\) as a function of driving frequency, as well as in \textbf{c,} the minimum \({\rm Min} |S_{\rm 21}|\) of \(|S_{\rm 21}|\) and \textbf{d,} the driving frequency \(\omega_{\rm min}\) at which this minimum occurs as a function of drive power.
    Data-points are rendered with blue dots and curve fits with solid lines.
    Each trace in \textbf{a} and \textbf{b} corresponds to a different drive power.
    At the top trace the drive power is \(P_{\rm in}= -135.23\) dBm, then this increases 1 dBm per trace, reaching \(P_{\rm in} = -118.23 \) dBm at the bottom trace.
    We offset every trace in \textbf{a} and \textbf{b} by -10 dB and -2 rad respectively to ease examination.
}
\end{figure}

The transformation between measured response $S_{21,\text{meas}}$ and fitted response $S_{21,\text{fit}}$ is thus given by
\begin{equation}
  \label{eq:fitted-response}
    (A+B\omega)e^{C+D\omega}S_{21,\text{fit}}(\omega) = S_{21,\text{meas}}\ ,
\end{equation}
where $A,B,C$ and $D$ are determined through a fit of a low-power response, where the nonlinearity does not come into play, and the response of the device alone $S_{21,\text{fit}}(\omega)$ is assumed to be
\begin{equation}
    S_{21,\text{fit}}(\omega) = 1-\frac{G}{i(\omega-F)+E}\ .
  \end{equation}

We then reduce the amount of noise as well as the superfluous number of frequency points in the data-set by replacing blocks of 10 successive frequency data-points by their average.
The treated data is notably used in the construction of Figs.~2,3.
The reduction in number of data-points also facilitates the fitting.
We were able to numerically compute $S_\text{21}$ over the 500 frequency points of the data-set in a minimization routine.
For each driving power of the data-set, the Python library QuTiP~\cite{johansson2012qutip,JOHANSSON20131234} was used to solve the Lindblad equation of Eq.~(\ref{eq:lindblad}) with \(n_{\rm th}=0\).
For each power, the absolute difference between the 500 (complex) numerical and experimental points constitute a first contribution to the minimized cost function.
The difference in the minimum of $|S_{\rm 21}|$, and the frequency at which $|S_\text{21}|$ is minimized, are also added to the cost-function each with a weight of 200 points.
The function is minimized using a modified Powell algorithm~\cite{2020SciPy-NMeth,powell1964efficient}, with five free-parameters: $\omega_\text{r}$, $\kappa_\text{int}$, $\kappa_\text{ext}$, $K$, and the attenuation that the signal outputted at the VNA experiences before reaching the device.
The attenuation is found to be $118.3$ dB, which is consistent with the physical attenuation installed at room-temperature and at the different stages of the dilution refrigerator.
The device parameters converge to 
$\omega_\text{r} = 2\pi\times 5.172$ GHz, $\kappa_\text{int}=2\pi\times 186$ kHz , $\kappa_\text{ext}=2\pi\times 2.12$ MHz, ${K=2\pi\times 80 \, {\rm kHz}}$.
Excellent agreement between this numerical computation based on the Lindblad equation and the experimental data is found for the complex transmission at all drive powers, as well as for the drive power dependence of both, the minimum value of $|S_\text{21}|$ and the driving frequency at which this minimum is reached, see Fig.~\ref{fig:lindblad-transmision-fits}.

The fitted parameters are confirmed by the measurement of a reference oscillator~\cite{yanai2019observation}, built using the same geometry as the device in Fig.~1 (and in the same fabrication run) but where the SQUID is replaced by a short circuit.
We simulate the resonance frequency of the reference oscillator using the finite-element software Sonnet, and compare it to the experimentally measured value.
The discrepancy between these allows us to determine the kinetic inductance of the 60 nm sheet of MoRe.
Using a sheet inductance of 1.575 pH/sq in Sonnet, the simulated resonance frequency matches the measured value.
We then add a lumped element inductor $L_\text{J}$ at the location of the SQUID in the simulation, and vary it to determine the value of the lumped element inductor $L$ and capacitor $C$, by fitting simulated resonance frequencies to $1/\sqrt{(L+L_\text{J})C}$.
The Josephson inductance $L_\text{J}$ is found when this simulated resonance matches the low-power resonance frequency measured in Fig.~2.
The circuit parameters obtained from this analysis are $L=2.93$ nH, $C=288$ fF and $L_\text{J} = 0.35$ nH. 
The resonance frequency and Kerr constant calculated from these parameters matches those determined by fitting the data with small deviations of 0.1\% and 2\% respectively.

Finally, we would like to acknowledge that further measurements could have been performed to validate both the system parameters and the physical mechanism of apparent non-linear damping.
The Kerr constant could be verified by strongly driving the system off-resonance and measuring the resulting Stark shift with a low-power probe.
The effective non-linear damping parameter $\gamma$ could be verified through the change in $S_\text{11}$ expected from a varied thermal occupation $n_\text{th}$. 
Here thermal occupation could be varied by injecting a known white noise or by increasing the base temperature of the dilution refrigerator.
Lastly, the SQUID tunability could be used to demonstrate an understanding of the apparent non-linear damping with a different Kerr constant and resonator frequency. 
Unfortunately, such measurements can no longer be carried out in a practical timescale: the device was fabricated for a different purpose and measured long before the analysis presented this work was performed and is no longer in working condition.

\section{Classical, noise-less solution and bifurcation point}
\label{sec:S_classical_solution}
Taking the expectation value of Eq.~(\ref{eq:langevin}), in the steady-state ($d\langle\hat a\rangle/dt=0$), yields
\begin{equation}
    i\Delta \langle\hat a\rangle -iK \langle\hat a^\dagger\hat a\hat a\rangle  +\frac{\kappa}{2} \langle\hat a\rangle=\epsilon\ .
\end{equation}
In the absence of (thermal and quantum) noise $\hat s = 0$, using the notation $ \langle\hat a\rangle = a$, one can simply write $\langle\hat a^\dagger\hat a\hat a\rangle = |a|^2 a$.
This yields the classical steady-state equation
\begin{equation}
    \left(i\Delta-iK |a|^2+\frac{\kappa}{2}\right)a  =\epsilon\ .
    \label{eq:classical_steady_state_equation}
\end{equation}
Note that $\Delta = \omega_\text{r}-\omega_\text{d}$ in the classical case, as the extra $-K$ in the definition of $\Delta$ in the quantum Langevin equation (Eq.~(\ref{eq:langevin})) is a consequence of the quantum fluctuations similar to the Lamb shift~\cite{gely2018nature}.
One can see this by rewriting the Kerr term $\hat a^\dagger\hat a ^\dagger\hat a\hat a$ using the commutation relations, which would yield a different expression for $\Delta$ in the quantum equation, but would not yield a different Kerr term in the classical steady-state equation.
To simulate nonlinear damping, we consider that the internal damping rate can depend on power $\kappa_\text{int}\rightarrow\kappa_\text{int}+\gamma|a|^2$ yielding
\begin{equation}
    \left(i\Delta-iK |a|^2+\frac{\kappa+\gamma|a|^2}{2}\right)a  =\epsilon\ .
    \label{eq:classical_steady_state_equation_nonlinear_damping}
\end{equation}

We solve this equation by computing the magnitude $|a|$ and phase $\varphi$ of the amplitude $a = |a|e^{i\varphi}$ separately.
An equation for $|a|$ is obtained by multiplying Eq.~(\ref{eq:classical_steady_state_equation}) by its conjugate, yielding
\begin{equation}
    \left(\frac{\gamma^2}{4}+K^2\right)|a|^6+\left(\frac{\gamma\kappa}{2}-2\Delta K\right)|a|^4+\left(\frac{\kappa^2}{4}+\Delta^2\right)|a|^2=\epsilon^2\ .
    \label{eq:nonlinear_damping_|a|2}
\end{equation}
We solve this equation by computing the eigenvalues of the polynomial's companion matrix~\cite{horn2012matrix} using Python~\cite{2020SciPy-NMeth}.
The phase is then given by
\begin{equation}
\varphi = \arctan \bigg( \frac{2(\Delta-K|a|^2)}{\kappa + \gamma|a|^2} \bigg) \, .
\end{equation}
Beyond a critical drive power there are drive frequencies for which Eq.~(\ref{eq:nonlinear_damping_|a|2}) possesses not only one but three real solutions.
In that case, the steady-state response of the oscillator will exhibit bistability as the drive frequency or detuning is varied.
The point (detuning) at which the onset of bistability takes place is known as a critical point (detuning).
The same applies to Eq.~(\ref{eq:nonlinear_damping_|a|2}) with \(\gamma = 0\), in which case the ensuing critical power \(P_{\rm c}=2 \sqrt{3}\hbar\omega_{\rm r}\kappa^{3}/(9|K|\kappa_{\rm ext})\simeq -122.2\) dBm is obtained through a simple stability analysis, see e.g. Appendix C of Ref.~\cite{brock2021nonlinearcharge}.
In what follows, we shall then use \(P_{\rm c}\) as an estimate of the actual critical power that follows with \(\gamma\neq 0\).
An examination of Figs.~\ref{fig:lindblad-transmision-fits}(a,b), suggests that this is a reasonable approximation.

Most of our measurements employ drive powers lower than \(P_{\rm c}\).
Furthermore, we observe that the phenomenon we are concerned with, namely, the reduction in the oscillator's amplitude as the drive power increases, manifests clearly in all those measurements.
Either a fully classical description of a steady-state in a nonlinear quantum system (the superconducting resonant circuit in our case) or that based on a linearization of quantum fluctuations around such a steady-state, are accurate far from a critical point.
However, both descriptions are known to fail in the vicinity of a critical point, see Ref.~\cite{degenfeld2015self} for more details on this regard.
As a consequence, we focus all the theoretical analysis we present in this work on a frequency response of the resonant circuit describable by a steady-state oscillator well below the bistability threshold.
To ensure this condition, we only consider drive powers \(P_{\rm in}\le -124\) dBm, further motivated by Sec.~\ref{sec:S_quantum_perturbation}.

\begin{figure}[htb]
  \begin{centering}
    \includegraphics[width=0.9\textwidth]{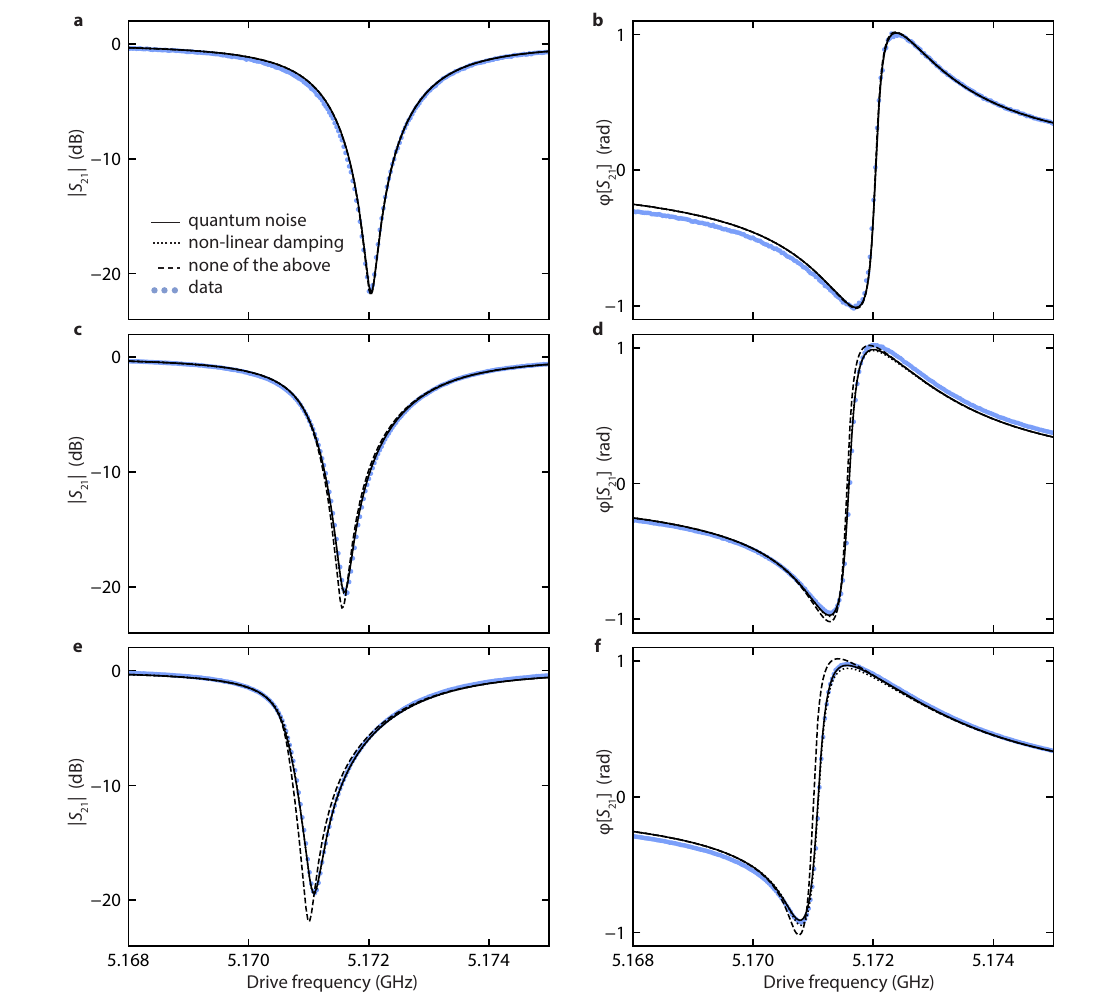}
    \par\end{centering}  
    \caption{\label{fig:S_fitted}
  \textbf{Quantum vs. classical descriptions: transmission coefficient.}
  Data-points (blue dots) of the transmission coefficient \(S_{\rm 21}\), processed following Sec.~\ref{sec:S_data_processing_fitting}, are compared to the best curve fits resulting from both a numerical solution of the Lindblad equation~(\ref{eq:lindblad}) (solid line), and a numerical solution of the classical steady-state equation~(\ref{eq:classical_steady_state_equation_nonlinear_damping}) with nonlinear damping (dotted line), following, respectively, the minimization routines described in Sec.~\ref{sec:S_data_processing_fitting} and Sec.~\ref{sec:S_classical_solution}.
  We also display the solution to the steady-state equation~(\ref{eq:classical_steady_state_equation}) of the linearly damped driven classical Kerr oscillator (dashed line). \textbf{a, c, e} The magnitude \(|S_{\rm 21}|\) and \textbf{b, d, f} phase \(\varphi[S_{\rm 21}]={\rm Arg}(S_{\rm 21})\) of the transmission coefficient are plotted as a function of the drive frequency.
Panels (a,b), (c,d) and (e,f) correspond to powers -135 dBm, -127 dBm and -124 dBm respectively; all of them associated with a steady-state oscillator below the bistability threshold.}
  \end{figure}

  \begin{figure}[htb]
\begin{centering}    
  \includegraphics[width=0.5\textwidth]{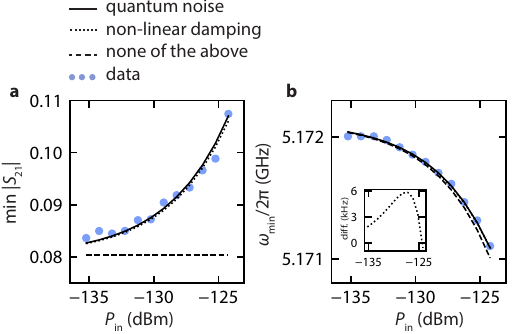}
\par\end{centering}  
\caption{\label{fig:S_fitted_min_fmin}
\textbf{Quantum vs. classical descriptions: minimum of $|S_\text{21}|$.}
\textbf{a}, The minimum \({\rm Min}|S_{\rm 21}|\) of transmission $|S_\text{21}|$ and \textbf{b}, the frequency $\omega_\text{min}$ which minimizes it, are plotted as a function of the drive power.
We obtain excellent agreement between experimental data (dots), the fit based on a numerical solution of the Lindblad master equation (solid line), as described in Sec.~\ref{sec:S_data_processing_fitting}, and the fit based on a numerical solution of the classical nonlinear damping model (dotted line), as described in Sec.~\ref{sec:S_classical_solution}. On the contrary, in \textbf{a} the minimum of transmission \(|1-\kappa_{\rm ext}/\kappa|\) that follows from the solution to the classical steady-state equation~(\ref{eq:classical_steady_state_equation}) with linear damping (dashed line) contrasts sharply with the experimental data.
In \textbf{b}, due to the overlap between the solid and dotted lines, the inset shows their difference as a function of drive power.
The drive power is bound to \(P_{\rm in}\le-124\) dBm, in which case the steady-state of the oscillator lies below the bistability threshold.
    }
\end{figure}  

Bearing in mind the above, we now search for the largest attainable amplitude for a given driving strength in Eq.~(\ref{eq:nonlinear_damping_|a|2}) by assuming a small deviation from the largest attainable amplitude without nonlinearities
\begin{equation}
    \begin{split}
        |a| &= \alpha(1+\delta)\ ,\\
        |a|^n &\simeq \alpha^{n}(1+n\delta)\\
    \end{split}
\end{equation}
with $\delta\ll 1$.
The amplitude $\alpha$ is the solution to
\begin{equation}
    -\frac{\kappa}{2}a+\epsilon=0\ ,
\end{equation}
the resonantly driven system, without nonlinearities.
Assuming a small deviation from $\alpha$ means we are considering the nonlinearities $K,\gamma$ as well as the detuning $\Delta$ to be perturbations, such that 
\begin{equation}
    K,\gamma,\Delta\ll \kappa\ .
    \label{eq:nonlinear_damping_approximation_K_gamma_Delta}
\end{equation}
Since $\Delta\simeq K\alpha^2$, we note that these approximations are only valid at the lower powers of our experimental data.
Injecting the perturbed expression for $|a|$ into Eq.~(\ref{eq:nonlinear_damping_|a|2}), we obtain an equation for $\delta$
\begin{equation}
    \left(\gamma^2/4+K^2\right)\alpha^6(1+6\delta)+\left(\gamma\kappa/2-2\Delta K\right)\alpha^4(1+4\delta)+\left(\kappa^2/4+\Delta^2\right)\alpha^2(1+2\delta)=\epsilon^2\ .
\end{equation}
Expanding the solution
\begin{equation}
    \delta=\frac{\epsilon^2-\left(\gamma^2/4+K^2\right)\alpha^6-\left(\gamma\kappa/2-2\Delta K\right)\alpha^4-\left(\kappa^2/4+\Delta^2\right)\alpha^2}{6\left(\gamma^2/4+K^2\right)\alpha^6+4\left(\gamma\kappa/2-2\Delta K\right)\alpha^4+2\left(\kappa^2/4+\Delta^2\right)\alpha^2}\ ,
\end{equation}
to second order in $K,\gamma,\Delta$ through the approximation of Eq.~(\ref{eq:nonlinear_damping_approximation_K_gamma_Delta}) yields
\begin{equation}
    \delta\simeq-\frac{\alpha^2\gamma}{\kappa}-\frac{4K^2\alpha^4-7\alpha^4\gamma^2-8K\alpha^2\Delta+4\Delta^2}{2\kappa^2}
\end{equation}
which is minimized for 
\begin{equation}
    \Delta = K\alpha^2
    \label{eq:S_Delta_nonlinear_solution}
\end{equation}
with a maximum (to leading order in $\gamma$)
\begin{equation}
    \delta = -\frac{\alpha^2\gamma}{\kappa}\ .
    \label{eq:S_delta_nonlinear_solution}
\end{equation}

In order to demonstrate that a classical nonlinear damping model also provides an adequate description of the data, we fit the data to a solution to Eq.~(\ref{eq:classical_steady_state_equation_nonlinear_damping}).
The data processing, and the construction of the cost function, is identical to that described in Sec.~\ref{sec:S_data_processing_fitting}.
The free-parameters, however, are different: we fix the internal and external damping, the oscillator frequency to $\omega_r-K$ and the attenuation to the values fitted with the quantum model.
The remaining free parameters of the fit are the nonlinear damping rate $\gamma$, and the Kerr nonlinearity $K$.
After convergence of the same modified Powell algorithm~\cite{powell1964efficient,2020SciPy-NMeth}, we obtain $\gamma = 2\pi\times5.02$ kHz and $K = 2\pi\times 78.3$ kHz (1.7 kHz lower than the value obtained by fitting the quantum model).
In Figs.~\ref{fig:S_fitted},\ref{fig:S_fitted_min_fmin}, we show that this model provides a reasonable fit to the data we address in our theoretical description (below the bistability threshold of the steady-state of our model oscillator).

\section{Quantum and noisy solution in the Gaussian regime}
\label{sec:S_quantum_perturbation}

Here we derive an approximate solution to the Langevin equation of Eq.~(\ref{eq:langevin}) in the presence of quantum and/or thermal noise.
The aim is to solve the equation for the ensemble average $\langle\hat{a}\rangle$ 
\begin{equation}
\frac{d }{d t} \langle \hat{a} \rangle = -[ i \Delta + \frac{\kappa}2] \langle \hat{a} \rangle + i  K \langle \hat{a}^\dagger\hat{a}^2\rangle + \epsilon\ ,
\label{eq:S_langevin_expectation_a_nonlinearized}
\end{equation}
in the steady-state and in a small neighborhood of the resonance.
In order to lighten our notation, we drop the time dependency in the equations and, unless it is explicitly stated otherwise, identify the ensuing dynamical variables with those that define the steady-state of the system.
Contrary to the noise-less case studied in the previous section, here $\langle\hat a^\dagger\hat a\hat a\rangle \ne |a|^2 a$.
This nonlinear term in Eq.~(\ref{eq:S_langevin_expectation_a_nonlinearized}) above leads to a Bogoliubov-Born-Green-Kirkwood-Yvon (BBGKY) like hierarchy of coupled differential equations, in which, e.g, the dynamics of the moment \(\left\langle \hat{a}^{n} \right\rangle\) of order \(n\) requires knowledge of the higher order moment \(\left\langle \hat{a}^{\dagger}a^{n+1} \right\rangle\), with \(n \in \mathbb{N}^{+}\) \cite{verstraelen2018gaussian}.
To perform a tractable analytical study of such infinite hierarchy of equations we shall truncate it appropriately.
We do so using a Gaussian state approximation which allows for closing the hierarchy, thus guaranteeing a physically meaningful steady-state solution \cite{degenfeld2015self,navarrete2014regularized,verstraelen2018gaussian}.
The resulting closed system of coupled differential equations involves only first order statistical moments (\textit{i.e.} ensemble averages) and second order statistical moments (covariances) of the system's canonical operators.
This approach, different to assuming a coherent state ansatz for the oscillator's quantum state, does not prescribe an operator's ensemble average to be entirely ruled by the classical dynamics of the system.
Instead, in this case an operator's ensemble average is treated as an unknown variable to be determined self-consistently with its covariances, thus including some information of the quantum dynamics of the system into that of the operator's ensemble averages.
Next, we present the conditions under which we may apply such an approach, followed by the description of a method for obtaining the ensemble average $\langle\hat{a}\rangle$ resulting from our Gaussian state ansatz.

\subsection{Assumptions and small quantities}

We assume that the thermal fluctuations, as well as the nonlinearity, have a small effect on the system.
This approximation comes in two forms.
First, we will assume the thermal fluctuations entering through the noise term $\hat s$ to be small through the condition
\begin{equation}
    n_\text{th}\ll |\langle\hat a\rangle|^2\ .
    \label{S_approximation_small_nth}
\end{equation}
Secondly, we will assume the anharmonicity to be small, through the condition
\begin{equation}
\begin{split}
    K\ll\kappa\ .
\end{split}
    \label{eq:S_approximation_small_K}
\end{equation}
In this regime, it is practical to express the equations using the deviation $\hat d$ from the steady-state expectation values of the operators involved in the dynamics.
For $\hat a$, this means we will write 
\begin{equation}
\hat{a} =\langle \hat{a} \rangle+\hat{d}\ .
\label{eq:d_definition}
\end{equation}
We will take $\hat d$ to be a small quantity through the assumption
\begin{equation}
	\label{eq:small_d_approximation}
	\begin{aligned}
		&\left|{\rm Re}\big[\langle\hat d^{\dagger n}\hat d^m\rangle\big]\right| \ll \left|{\rm Re}\big[\langle \hat{a}^{\dagger n} \hat{a}^{m} \rangle - \langle\hat d^{\dagger n}\hat d^m\rangle\big]\right|
	\, \text{and} \, \left|{\rm Im}\big[\langle\hat d^{\dagger n}\hat d^m\rangle\big]\right| \ll \left|{\rm Im}\big[\langle \hat{a}^{\dagger n} \hat{a}^{m} \rangle - \langle\hat d^{\dagger n}\hat d^m\rangle\big]\right| \, \text{if} \; n+m>2 \, ,
	\end{aligned}
\end{equation}
such that we will neglect $\langle\hat d^{\dagger n}\hat d^m\rangle$ with respect to $\langle \hat{a}^{\dagger n} \hat{a}^{m} \rangle - \langle\hat d^{\dagger n}\hat d^m\rangle$ and thus use $\langle \hat{d}^{\dagger n} \hat{d}^{m} \rangle \simeq 0$ whenever $n+m>2$.
This second order ladder approximation amounts to considering that the steady-state is well described by a Gaussian state.
We note that for a Gaussian state $\langle(\hat d^{\dagger}\hat d)^m\rangle$ scales with $n_\text{th}^m$.
Therefore, considering that the thermal fluctuations are small $n_\text{th}\ll|\langle \hat{a} \rangle|^2$ is necessary for Eq.~(\ref{eq:small_d_approximation}) to hold.
The assumption~(\ref{eq:small_d_approximation}) enables a linearization of the steady-state equations of the ensemble averages and covariances of the oscillator's canonical operators that fully characterize the resulting Gaussian steady-state.
As shown for a similar paradigm in Ref.~\cite{navarrete2014regularized}, given a limit of weak anharmonicity, c.f.~Eq.~(\ref{eq:S_approximation_small_K}), and far from a multistability threshold, this linearization leads to an analytical steady-state solution in good agreement with numerical calculations.
This approach is then applicable for our system below the bistability threshold.
Although lacking the precision of other methods~\cite{drummond1980quantum,kheruntsyan1999wigner}, it provides a faithful description of our experimental observations, with the additional benefit that quantities of interest (such as averages and covariances of canonical operators) can be expressed more simply.

\subsection{Preliminary results}

We first derive expressions for useful expectation values.
Injecting Eq.~(\ref{eq:d_definition}) in $\langle\hat{a}^\dagger\hat{a}\rangle$ for example yields 
\begin{align}
    \langle\hat{a}^\dagger\hat{a}\rangle &= \left\langle (\langle \hat a\rangle^*+\hat d^\dagger )(\langle \hat a\rangle+\hat d)\right\rangle\\
    &= \langle \hat a\rangle^*\langle \hat a\rangle+\langle \hat a\rangle^*\langle\hat d\rangle+\langle \hat a\rangle\langle\hat d^\dagger\rangle + \langle\hat d^\dagger\hat d\rangle\\
    &= \langle \hat a\rangle^*\langle \hat a\rangle+\langle \hat d^\dagger\hat d\rangle\ ,
    \label{eq:S_adag_a_from_ddag_d}
\end{align}
where we have used the definition of $\hat d$ to obtain $\langle\hat d\rangle=\langle\hat d^\dagger\rangle=0$.
We proceed similarly to obtain
\begin{align}
    \langle\hat{a}^2\rangle 
    &= \langle \hat a\rangle^2+\langle\hat d^2\rangle\ ,
    \label{eq:S_a2_from_d2}
\end{align}
and by invoking the approximation of Eq.~(\ref{eq:small_d_approximation}), we also have
\begin{equation}
\langle \hat{a}^\dagger\hat{a}^2\rangle \simeq 2\langle \hat{d}^\dagger\hat{d}\rangle\langle \hat{a}\rangle + \langle \hat{a}^{\dagger}\rangle\langle \hat{d}^2\rangle + |\langle \hat{a}\rangle|^2\langle \hat{a}\rangle\ ,
\label{eq:adag_a2_approximated}
\end{equation}
\begin{align}
\langle \hat{a}^\dagger\hat{a}^3\rangle 
&\simeq |\langle\hat{a}\rangle|^2\langle\hat{a}\rangle^2+3|\langle\hat{a}\rangle|^2\langle\hat{d^2}\rangle +3 \langle\hat{a}\rangle^2 \langle\hat{d}^\dagger\hat d\rangle\ .
\label{eq:S_adag_a3_from_d2}
\end{align}

\subsection{Reformulation of the problem}

By injecting Eq.~(\ref{eq:adag_a2_approximated}) in Eq.~(\ref{eq:S_langevin_expectation_a_nonlinearized}), we rewrite the equation for $\langle \hat{a} \rangle$ as
\begin{align}
& \frac{d }{d t} \langle \hat{a} \rangle \simeq -\big[ i (\Delta-2K\langle\hat{d}^\dagger\hat{d}\rangle -K|\langle \hat{a}\rangle|^2)+\kappa/2\big]\langle \hat{a} \rangle + i  K \langle \hat{a}^{\dagger}\rangle\langle\hat{d}^2\rangle + \epsilon\text{.}
\label{eq:S_langevin_expectation_a}
\end{align}
A steady-state solution for $\langle \hat{a} \rangle$ thus requires knowledge of $\langle\hat{d}^\dagger\hat{d}\rangle$ and $\langle \hat{d}^2\rangle$.
These expectation values will be determined by deriving the equation of motion of $\hat{d}^\dagger\hat{d}$ and $\hat{d}^2$.

\subsection{Equation of motion for $\langle\hat{d}^\dagger\hat{d}\rangle$}

From Eq.~(\ref{eq:S_adag_a_from_ddag_d}), we have $\langle\hat{d}^\dagger\hat{d}\rangle = \langle\hat a^\dagger\hat a\rangle -\langle \hat a\rangle^*\langle \hat a\rangle$
such that
\begin{equation}
    \frac{d }{d t}\langle\hat{d}^\dagger\hat{d}\rangle = \frac{d }{d t}\langle\hat{a}^\dagger\hat{a}\rangle - \langle\hat{a}\rangle^*\frac{d }{d t}\langle\hat{a}\rangle- \langle\hat{a}\rangle\left(\frac{d }{d t}\langle\hat{a}\rangle\right)^*\ .
\end{equation}
We should thus search for the equation of motion for $\langle\hat a^\dagger\hat a\rangle$.
Utilizing the Langevin equation of Eq.~(\ref{eq:langevin}), we first obtain an equation for $\hat a^\dagger\hat a$
\begin{align}
    \frac {d}{dt}\left(\hat{a}^\dagger\hat{a}\right) &= \left(\frac{d}{dt}\hat a^\dagger\right)\hat a+\hat a^\dagger\left(\frac{d}{dt}\hat a\right)\\
    & =- \kappa\hat{a}^\dagger\hat{a} + \epsilon (\hat{a}+\hat{a}^{\dagger}) - \sqrt{\kappa}(\hat{a}^{\dagger} \hat{s}  + \hat{s}^{\dagger}  \hat{a} )\ .
\end{align}
When taking the expectation value of this equation, we follow Ref.~\cite{gardiner2015quantum} to treat the noise terms.
For $\hat s$ given by quantum noise, with $\langle\hat s\rangle=0$ and Eqs.~(\ref{eq:s_correlations}), then given $\hat A$ an arbitrary system operator, we have
\begin{equation}
    \langle \hat A(t)\hat s(t)\rangle = n_\text{th}\frac{\sqrt{\kappa}}2\ \langle[\hat A(t),\hat a(t)]\rangle\ ,
    \label{eq:As}
\end{equation}
\begin{equation}
    \langle\hat s^\dagger(t) \hat A(t)\rangle = n_\text{th}\frac{\sqrt{\kappa}}2\ \langle[\hat a^\dagger(t),\hat A(t)]\rangle\ ,
    \label{eq:sdA}
\end{equation}
such that $\langle\hat{a}^{\dagger} \hat{s}  + \hat{s}^{\dagger}  \hat{a} \rangle = -\sqrt{\kappa}n_\text{th}$.
Using Eq.~(\ref{eq:S_adag_a_from_ddag_d}) again to rewrite $\langle\hat a^\dagger\hat a\rangle$ we have
\begin{align}
  \label{eq:motion-adag-a}
    \frac {d}{dt}\langle\hat a^\dagger\hat a\rangle &=-\kappa(|\langle \hat a\rangle|^2+\langle \hat d^\dagger\hat d\rangle) + \epsilon (\langle\hat{a}\rangle+\langle\hat{a}^{\dagger}\rangle)+\kappa n_\text{th}\ .
\end{align}
Using the equation of motion for $\langle\hat a\rangle$ of Eq.~(\ref{eq:S_langevin_expectation_a}), we finally obtain
\begin{align}
\frac{d }{d t}\langle\hat{d}^\dagger\hat{d}\rangle &= \frac{d }{d t}\langle\hat{a}^\dagger\hat{a}\rangle - \langle\hat{a}\rangle^*\frac{d }{d t}\langle\hat{a}\rangle- \langle\hat{a}\rangle\left(\frac{d }{d t}\langle\hat{a}\rangle\right)^* \\
&\simeq - \kappa\langle\hat{d}^\dagger\hat{d}\rangle +\kappa n_\text{th} + i K \left(\langle \hat{d}^{\dagger 2}\rangle\langle \hat{a}\rangle^2 - \langle \hat{a}^{\dagger}\rangle^2\langle \hat{d}^2\rangle\right)\ .
\label{eq:S_langevin_expectation_ddagd}
\end{align}

\subsection{Equation of motion for $\langle \hat{d}^2\rangle$}

From Eq.~(\ref{eq:S_a2_from_d2}) we have $\langle\hat{d}^2\rangle = \langle\hat a^2\rangle -\langle \hat a\rangle^2$, such that the equation of motion for  $\langle\hat{d}^2\rangle$ writes
\begin{equation}
     \frac{d }{d t}\langle\hat{d}^{2}\rangle =\frac{d }{d t}\langle\hat{a}^{2}\rangle - 2\langle\hat{a}\rangle\left(\frac{d }{d t}\langle\hat{a}\rangle\right)\ .
\end{equation}
We should thus search for the equation of motion for $\langle\hat a^2\rangle$.
We start by writing the equation of motion for $\hat a^2$ as 
\begin{equation}
\begin{split}
    \frac{d}{dt}\hat a^2 &= \hat a\left(\frac{d}{dt}\hat a\right)+\left(\frac{d}{dt}\hat a\right)\hat a\\
    &=-i\left(2\Delta-K\{\hat a^\dagger,\hat a\}\right)\hat a^2-\kappa \hat a^2+2\epsilon\hat a-\sqrt{\kappa}\left(\hat s\hat a+\hat a\hat s\right)\ .
\end{split}
\end{equation}
If quantum noise is taken into account, then we have $\{\hat a^\dagger,\hat a\} = \color{blue}1\color{black}+2\hat a^\dagger\hat a$ rather than $\{ a^*, a\} = 2 a^* a$ without.
To keep track of the quantum noise, we have colored this term in blue throughout the rest of the derivation.
%
When taking the expectation values, we treat the noise terms in $\hat s$ following Eq.~(\ref{eq:As}), and an additional result from Ref.~\cite{gardiner2015quantum}
\begin{equation}
    \langle \hat s(t)\hat A(t)\rangle = (n_\text{th}+1)\frac{\sqrt{\kappa}}2\ \langle[\hat A(t),\hat a(t)]\rangle\ ,
    \label{eq:sA}
\end{equation}
resulting in $\langle\hat s\hat a+\hat a\hat s\rangle=0$.
By rewriting $\langle\hat a^2\rangle$ using Eq.~(\ref{eq:S_a2_from_d2}) and $\langle\hat a^\dagger \hat a^3\rangle$ using Eq.~(\ref{eq:S_adag_a3_from_d2}), we get
\begin{equation}
    \frac{d}{dt} \langle \hat a^2\rangle \simeq 2iK\left(|\langle\hat{a}\rangle|^2\langle\hat{a}\rangle^2+3|\langle\hat{a}\rangle|^2\langle\hat{d^2}\rangle +3 \langle\hat{a}\rangle^2 \langle\hat{d}^\dagger\hat d\rangle\right)-\left(2i\Delta-\color{blue}iK\color{black}+\kappa\right)\left(\langle \hat a\rangle^2+\langle\hat d^2\rangle\right) +2\epsilon\langle \hat a\rangle\ .
\end{equation}
Using this equation as well as the equation of motion for $\langle\hat a\rangle$ of Eq.~(\ref{eq:S_langevin_expectation_a}), we finally obtain
\begin{align}
 \frac{d }{d t}\langle\hat{d}^{2}\rangle &=\frac{d }{d t}\langle\hat{a}^{2}\rangle - 2\langle\hat{a}\rangle\left(\frac{d }{d t}\langle\hat{a}\rangle\right)\\
&\simeq-2\left(i\left(\Delta-2K(|\langle \hat{a}\rangle|^2+\color{blue}\frac14\color{black} )\right)+\kappa/2\right)\langle\hat{d}^2\rangle+ i  2K\langle \hat{a}\rangle^2\left(\langle \hat{d}^\dagger\hat{d}\rangle+\color{blue}\frac12\color{black}\right)\ .
\label{eq:S_langevin_expectation_d2}
\end{align}

\subsection{Steady-state solution}

We obtain a steady-state solution for the equations of motion (\ref{eq:S_langevin_expectation_a},\ref{eq:S_langevin_expectation_ddagd},\ref{eq:S_langevin_expectation_d2}) by equating all time derivatives to zero.
We first solve for the covariances to get
\begin{align}
  \langle \hat{d}^\dagger\hat{d}\rangle &\simeq \frac{n_\text{th} + \color{blue}\frac{1}{2}\color{black}\frac{4K^2|{\langle \hat{a}\rangle}|^4}{4\Omega^2+\kappa^2}}{1- \frac{4K^2|{\langle \hat{a}\rangle}|^4}{4\Omega^2+\kappa^2}}=\frac{n_\text{th} + \color{blue}\frac{1}{2}\color{black}}{1- \frac{4K^2|{\langle \hat{a}\rangle}|^4/\kappa^2}{4\Omega^2/\kappa^2+1}} - \frac{1}{2},
\label{eq:ddagcov}
  \\
  {\langle \hat{d}^2\rangle} &\simeq  i \frac{K{\langle \hat{a}\rangle}^2}{ i  \Omega + \kappa/2}\left(\langle \hat{d}^\dagger\hat{d}\rangle +\color{blue}\frac{1}{2}\color{black}\right) =  \frac{4K}{\kappa}\langle \hat{a}\rangle^2\frac{n_\text{th} + \color{blue}\frac{1}{2}\color{black}}{4\Omega^2/\kappa^2- 4K^2|{\langle \hat{a}\rangle}|^4/\kappa^{2} + 1}(\Omega/\kappa + i/2)\,
\label{eq:dcov}
\end{align}
where $\Omega = \Delta - K(2|\langle \hat{a}\rangle|^2 + \color{blue}1/2\color{black})$.
By injecting our expressions for $\langle \hat{d}^\dagger\hat d\rangle,{\langle \hat{d}^2\rangle}$ (c.~f. Eqs.~(\ref{eq:ddagcov}) and (\ref{eq:dcov}), respectively) in Eq.~(\ref{eq:S_langevin_expectation_a}) and after some algebra, we find that, in the steady-state, the amplitude is ruled by the following equation

\begin{align}
  & i\left[\left(\Delta-\frac{K}{2}\right)\left(1 - \frac{\gamma|\langle \hat{a} \rangle|^2/\kappa}{4\Omega^2/\kappa^2- 4K^2|{\langle \hat{a}\rangle}|^4/\kappa^{2}+1}\right) +\frac{3K}{2}-K|\langle \hat{a} \rangle|^2- 2K\left(n_\mathrm{th} + \color{blue}\frac{1}{2}\color{black}\right) \right]\langle \hat{a} \rangle \nonumber
  \\
  &+\frac{\kappa}{2}\left[1 + \frac{\gamma|\langle \hat{a} \rangle|^2/\kappa}{4\Omega^2/\kappa^2- 4K^2|{\langle \hat{a}\rangle}|^4/\kappa^{2} + 1}\right]\langle \hat{a}\rangle \simeq \epsilon\ .
\label{eq:steady-state-amp}
\end{align}
Equation~(\ref{eq:steady-state-amp}) shows that the interplay of the oscillator's nonlinearity and the noise gives rise, among other effects, to a nonlinear broadening of the oscillator's steady-state response, the strength of which is given by
\begin{equation}
    \gamma=\frac{4K^2}{\kappa}\left(n_\text{th}+\color{blue}\frac{1}{2}\color{black}\right)\ .
    \label{eq:S_gamma_n}
    \end{equation}
    These effects resulting from the interplay of the noise and the oscillator's nonlinearity manifest most prominently as we approach to a resonance scenario, for if, on the contrary \(|\Delta/\kappa|\to\infty\), the distribution \({1/(4\Omega^2/\kappa^2- 4K^2|{\langle \hat{a}\rangle}|^4/\kappa^{2} + 1)\to 0}\).
    Moreover, for a sufficiently weak anharmonicity, in a narrow frequency window enclosing the resonance and on resonance itself, we shall see next that \({1/(4\Omega^2/\kappa^2- 4K^2|{\langle \hat{a}\rangle}|^4/\kappa^{2} + 1) \simeq 1}\).
      In that case, a nonlinear broadening of the resonance is the dominant effect that stems from the presence of noise in the system.
      Indeed, in order to be able to approximate the value of the aforementioned distribution by 1 it is necessary that \(|4\Omega^2/\kappa^2- 4K^2|{\langle \hat{a}\rangle}|^4/\kappa^{2}| \ll 1\).
      The solutions to this inequality depend on the value of \(K|{\langle \hat{a}\rangle}|^2/\kappa\).
      That is, the solution to the inequality is given by the sets
      \begin{equation}
        \label{eq:inequality1} -\sqrt{\frac{K^2|\langle\hat{a}\rangle|^{4}}{\kappa^{2}}+\frac{1}{4}}\ll\frac{\Omega}{\kappa}\ll-\sqrt{\frac{K^2|\langle\hat{a}\rangle|^{4}}{\kappa^{2}}-\frac{1}{4}} \, \text{ and }\, \sqrt{\frac{K^2|\langle\hat{a}\rangle|^{4}}{\kappa^{2}}-\frac{1}{4}}\ll \frac{\Omega}{\kappa}\ll \sqrt{\frac{K^2|\langle\hat{a}\rangle|^{4}}{\kappa^{2}}+\frac{1}{4}}\, \text{ if }\, \frac{K|\langle\hat{a}\rangle|^{2}}{\kappa} \geq 1/2\, ,
      \end{equation}
      and by the set
      \begin{align}
        \label{eq:inequality}
        -\sqrt{\frac{K^2|\langle \hat{a}\rangle|^{4}}{\kappa^{2}}+\frac{1}{4}} \ll \frac{\Omega}{\kappa}\ll\sqrt{\frac{K^2|\langle\hat{a}\rangle|^{4}}{\kappa^{2}}+\frac{1}{4}}\, \text{ if }\, 0<\frac{K|{\langle \hat{a}\rangle}|^2}{\kappa}<\frac{1}{2} \ .
      \end{align}
      Using the oscillator's parameters measured in our experiment and considering either \(n_{\rm th}=0\),  \(n_{\rm th} = 0.05\) (the upper bound for the average number of thermal photons in our experiment as estimated in Sec.~\ref{sec:experimental-setup}) or \(n_{\rm th} = 1/2\), a numerical solution of the master equation (\ref{eq:lindblad}) reveals that our anharmonicity is weak enough so as to guarantee \(K|\langle \hat{a}\rangle|^{2}/\kappa \lesssim 0.48\) for every detuning, \(-4 {\rm MHz} \lesssim \Delta/(2\pi) \lesssim 2.92 {\rm MHz}\), and all the drive powers \(P_{\rm in}\leq -124\) dBm addressed in our analysis.
      Drive powers above \(P_{\rm in} = -124\) dBm but still close to the critical power (at about \(P_{\rm c}\simeq -122.2\) dBm) that sets the onset of a bistable steady-state, not only break down the condition \(K|\langle \hat{a}\rangle|^{2}/\kappa < 1/2\), but lead to a solution that starts satisfying more narrowly the necessary assumption~(\ref{eq:small_d_approximation}).
Let us then focus on this latter regime for which \(0<K|{\langle \hat{a}\rangle}|^2/\kappa<1/2\) is fulfilled.
For a given driving strength resonance is attained with the detuning \(\Delta_{\star}\) for which the magnitude \(\lvert\langle \hat{a}\rangle\rvert\) of the steady-state amplitude reaches its maximum \(\lvert\langle \hat{a}_{\star}\rangle\rvert\).
\begin{figure}[htb]
  \begin{centering}
\includegraphics[width= 0.64\linewidth]{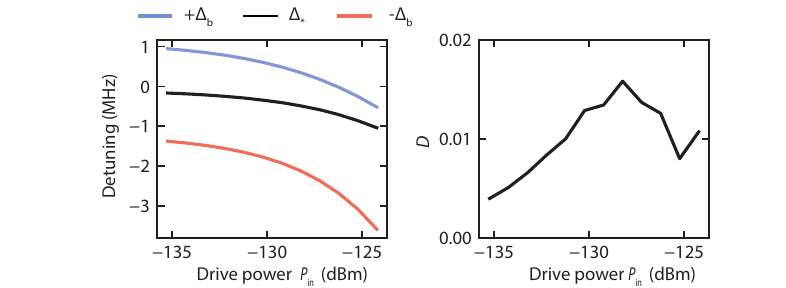} 
\par\end{centering}
\caption{\label{fig:negligible-distribution} 
\textbf{Conditions leading to a standard nonlinear broadening, proportional to \(|\langle \hat{a}\rangle_{\star}|^{2}\), of the resonance of a damped driven quantum Kerr oscillator close to a Gaussian state.}
In a regime complying with \(0 < K|\langle \hat{a}\rangle|^{2}/\kappa< 1/2\) for every applied input power, (a) the on resonance detuning, \(\Delta_{\star}\), lies in between the detuning bounds \(\pm \Delta_{\rm b}\) that derive from the frequency bounds \(\pm \Omega_{\rm b}\) delimiting the on resonance equivalent of the set defined in Eq.~(\ref{eq:inequality}) and, at the same time (b) both \(\Omega_{\star} = \Delta_{\star} - K(2|\langle \hat{a}\rangle_{\star}|^{2} +1/2)\) and the magnitude of the oscillator's amplitude maximum \(|\langle \hat{a}\rangle_{\star}|\) are such that the quantity \(D= |{[4\Omega_{\star}^2/\kappa^2- 4K^2|\langle \hat{a}\rangle_{\star}|^4/\kappa^{2} + 1]^{-1}}-1|\) respects our approximation $D\ll 1$.
All the plotted quantities are obtained from a numerical solution of the Lindblad eqation~(\ref{eq:lindblad}) with parameter values as extracted from the experimental measurements and \(n_{\rm th}=1/2\).
}
\end{figure}
Fig.~\ref{fig:negligible-distribution}(a) shows the dependence on \(P_{\rm in}\) of this resonance detuning \(\Delta_{\star}\) and the detuning bounds \({\pm\Delta_{\rm b} = K(2|\langle \hat{a}\rangle_{\star}|^2 + \color{blue}1/2\color{black})\pm \Omega_{\rm b}}\) that result from the frequency bounds \({\pm\Omega_{\rm b}/\kappa=\pm \sqrt{K^2|\langle \hat{a}\rangle_{\star}|^{4}/\kappa^{2}+1/4}} \) corresponding to the on resonance version of the set defined in Eq.~(\ref{eq:inequality}).
Note that knowledge of a maximum \(\lvert\langle \hat{a}\rangle_{\star}\rvert\) as computed through a numerical solution of the master equation (\ref{eq:lindblad}) enables us to determine both, its corresponding detuning \(\Delta_{\star}\) as well as the ensuing detuning bounds \(\pm\Delta_{\rm b}\).
That is how we obtain each detuning value we plot in Fig.~\ref{fig:negligible-distribution}(a), using in the numerical computation \(n_{\rm th}=1/2\) and the same parameter values we extract from the experimental data.
Clearly, we observe that \(-\Delta_{\rm b}<\Delta_{\star}<\Delta_{\rm b}\), i.e., that \((-\Delta_{\rm b},\Delta_{\rm b})\) defines an interval of detunings comprising the resonance.
In Fig.~\ref{fig:negligible-distribution}(b), we check the validity of the assumption \(-\Delta_{\rm b}\ll \Delta_{\star} \ll \Delta_{\rm b}\) or, more explicitly, the relative error of the approximation \({1/(4\Omega_{\star}^2/\kappa^2- 4K^2|{\langle \hat{a}\rangle}_{\star}|^4/\kappa^{2} + 1) \simeq 1}\) as a function of \(P_{\rm in}\), where \(\Omega_{\star} = \Delta_{\star} - K(2|\langle \hat{a}\rangle_{\star}|^2 + \color{blue}1/2\color{black}) \).
The curve shows that the deviation of the distribution from the target value 1 remains lower than a \(2\%\) for the entire range of input powers used in our theoretical analysis.
The same curve as computed using either \(n_{\rm th} = 0\) or \(n_{\rm th} = 0.05\) shows an error that stays below a \(5\%\) for input powers \(P_{\rm in} \leq -126\) dBm, after which the error increases up to a \(10\%\) for \(P_{\rm in}=-124\) dBm.
As we anticipated above, we may then conclude that Eq.~(\ref{eq:inequality}) is rather well fulfilled on resonance. 
Eq.~(\ref{eq:inequality}) can be equally satisfied for an arbitrarily small range of detunings \({\Delta_{\star}- \vartheta\leq \Delta\leq \Delta_{\star}+ \vartheta}\) around the resonance, with \(\vartheta\) a real constant such that \(0<\vartheta\ll|\Delta_{\star}|\).
Thus, in order to evaluate the oscillator's steady-state amplitude near resonance, we constrain our analysis to such range of detunings and set \({1/(4\Omega^2/\kappa^2- 4K^2|{\langle \hat{a}\rangle}|^4/\kappa^{2} + 1) \simeq 1}\).
This allows us to approximate the equation for the steady-state amplitude as
%
\begin{align}
  \label{eq:steady-state-amp-reduced}
    &i \left[\left(\Delta-\frac{K}{2}\right)\left(1- \frac{\gamma|\langle \hat{a}\rangle|^2}{\kappa}\right) +\frac{3K}{2}-K|\langle \hat{a} \rangle|^2- 2K\left(n_{\rm th} + \color{blue}\frac{1}{2}\color{black}\right)\right]\langle \hat{a} \rangle +\frac{\kappa}{2}\left[1 + \frac{\gamma|\langle \hat{a}\rangle|^2}{\kappa}\right]\langle \hat{a}\rangle \simeq \epsilon \, .
\end{align}
%
We may now associate the resonance scenario with the approximated value of the maximum of the steady-state amplitude's magnitude that results from this Eq.~(\ref{eq:steady-state-amp-reduced}).
We achieve this with the detuning ${\Delta=K/2 -[3K/2 - K|\langle \hat{a}\rangle_{\star}|^2- 2K(n_\mathrm{th} + \color{blue}1/2\color{black})][1-\gamma|\langle \hat{a}\rangle_{\star}|^2/\kappa]^{-1}}$ that cancels the first bracket in the left hand side of Eq.~(\ref{eq:steady-state-amp-reduced}), such that Eq.~(\ref{eq:steady-state-amp-reduced}) itself simplifies to \([1 + \gamma|\langle \hat{a}\rangle_{\star}|^2/\kappa]\langle \hat{a}\rangle_{\star} \simeq \alpha\), an equation that showcases the same form as the on resonance version of the steady-state equation~(\ref{eq:classical_steady_state_equation_nonlinear_damping}) of the nonlinearly damped driven classical oscillator introduced in Sec.~\ref{sec:S_classical_solution}.
Using Cardano's method we solve the equation \([1 + \gamma|\langle \hat{a}\rangle_{\star}|^2/\kappa]^{2}|\langle \hat{a}\rangle_{\star}|^2 \simeq \alpha^{2}\) for \(|\langle \hat{a}\rangle_{\star}|^2\).
The positive square root of such solution provides us with the approximated value of the maximum \(|\langle \hat{a}\rangle_{\star}|\) that we seek, which reads
\begin{equation}
\begin{split}
  &|\langle\hat a\rangle_{\star}|\simeq \left[\sqrt[3]{\frac{\alpha^{2}}{2\varepsilon^{2}}+\frac{1}{27\varepsilon^{3}}+\sqrt{\left(\frac{\alpha^{2}}{2\varepsilon^{2}}+\frac{1}{27\varepsilon^{3}}\right)^2-\frac{1}{729\varepsilon^6}}}+\sqrt[3]{\frac{\alpha^{2}}{2\varepsilon^{2}}+\frac{1}{27\varepsilon^{3}}-\sqrt{\left(\frac{\alpha^{2}}{2\varepsilon^{2}}+\frac{1}{27\varepsilon^{3}}\right)^2-\frac{1}{729\varepsilon^6}}}-\frac{2}{3\varepsilon}\right]^{1/2} \ ,\\
  &\varepsilon = \frac{\gamma}{\kappa}=\frac{4K^2}{\kappa^2}\left(n_\text{th}+\color{blue}\frac{1}{2}\color{black}\right) \ ,
    \label{eq:analytical_a_full}
\end{split}
\end{equation}
in good agreement with the data (see Fig.~\ref{fig:S_a_analytical}).

\begin{figure}[htb]
  \begin{centering}
    \includegraphics[width=0.5\textwidth]{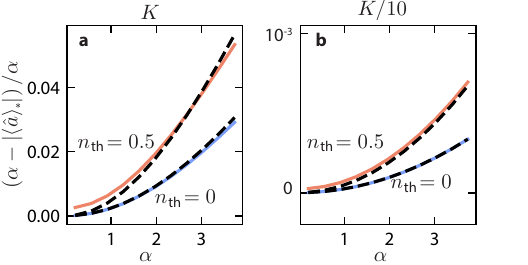}
    \par\end{centering}
\caption{
    \textbf{Test of the analytical model for $|\langle\hat a\rangle_{\star}|$.}
    Numerical calculations of $|\langle\hat a\rangle_{\star}|$ (full curves) are compared to the analytical expressions of Eq.~(\ref{eq:analytical_a_full}) (dashed curves).
    In panel \textbf{a} we use the same regime of parameters as the measured device, notably the same Kerr parameter $K$, with both no thermal population $n_\text{th}=0$ and half a quantum of thermal population $n_\text{th}=1/2$.
    In panel \textbf{b}, we show that the accuracy of the analytical model increases as we decrease $K$ by a factor 10.
    For the calculations with thermal population, good agreement is found in the regime $n_\text{th}\ll|\langle \hat{a} \rangle|^2$ covered by our assumptions.
    }
\label{fig:S_a_analytical}
\end{figure}


Similarly to Sec.~\ref{sec:S_classical_solution}, to ease the subsequent analysis and acquire a clearer insight of the behaviour of the oscillator's resonant amplitude, we derive next a simpler approximated solution to Eq.~(\ref{eq:steady-state-amp-reduced}) based on perturbation theory.
Given that \(K|{\langle \hat{a}\rangle}|^2/\kappa<1/2\), if we further take into account our original assumptions of weak anharmonicity and strength of thermal fluctuations, c.~f. (\ref{eq:S_approximation_small_K}) and (\ref{S_approximation_small_nth}), respectively, we find that \(\varepsilon \ll 1\).
The solution of the oscillator's steady-state amplitude that we seek comes from a series expansion \(\langle \hat{a} \rangle = \sum_{l=0}\varepsilon^l \langle \hat{a}\rangle_l\) in powers of the small parameter \(\varepsilon = \gamma/ \kappa\).
For our simple description of the resonance, we content ourselves with the regime for which the first order correction in \(\varepsilon\) is dominant.
This leaves us with the expansions \(\langle\hat{a}\rangle \simeq \langle\hat{a}\rangle_0 + \varepsilon \langle\hat{a}\rangle_1\) and \(\lvert\langle\hat{a}\rangle\rvert^{2} \simeq \lvert\langle\hat{a}\rangle_{0}\rvert^{2}+\varepsilon[\langle\hat{a}\rangle_{0}^{*}\langle\hat{a}\rangle_{1} + \langle\hat{a}\rangle_{1}^{*}\langle\hat{a}\rangle_{0}]\).
Likewise, we expand the detuning in the form \(\Delta = \Delta_{0} + \varepsilon \Delta_{1}\).
By replacing these expansions into Eq.~(\ref{eq:steady-state-amp-reduced}), neglecting terms of order higher than \(\varepsilon^{1}\) and grouping terms of the same order in \(\varepsilon\), we obtain the following set of equations
\begin{align}
\label{eq:steady-state-pert-zero}
  &i \left[\Delta_{0}+K-K|\langle \hat{a} \rangle_{0}|^2- 2K\left(n_{\rm th} + \color{blue}\frac{1}{2}\color{black}\right)\right]\langle \hat{a} \rangle_{0} +\frac{\kappa}{2}\langle \hat{a}\rangle_{0} = \epsilon \, ,
  \\
  &i\left[\Delta_{1}-|\langle \hat{a}\rangle_{0}|^{2}\Delta_{0}+\frac{K}{2}|\langle \hat{a}\rangle_{0}|^{2}-K(\langle\hat{a}\rangle_{0}^{*}\langle\hat{a}\rangle_{1} + \langle\hat{a}\rangle_{1}^{*}\langle\hat{a}\rangle_{0})\right]\langle \hat{a}\rangle_{0}\nonumber
  \\
  \label{eq:steady-state-pert-one}
    &+i \left[\Delta_{0}+K-K|\langle \hat{a} \rangle_{0}|^2- 2K\left(n_{\rm th} + \color{blue}\frac{1}{2}\color{black}\right)\right]\langle \hat{a}\rangle_{1}+\frac{\kappa}{2}\left[\langle \hat{a}\rangle_{1} + |\langle \hat{a}\rangle_{0}|^{2}\right]\simeq 0 \, .
\end{align}
In the resonance scenario described by the set of equations above, the detuning must be such that it cancels each of the brackets multiplied by the imaginary unit in the left hand side of Eqs.~(\ref{eq:steady-state-pert-zero})~and~(\ref{eq:steady-state-pert-one}).
Solving the resulting equations for the corresponding resonant amplitude yields \(\langle \hat{a}\rangle_{\star} \simeq \langle \hat{a}\rangle_{\star,0}+\varepsilon\langle \hat{a}\rangle_{\star,1}\) with \(\langle \hat{a}\rangle_{\star,0}=\alpha\) and \({\langle \hat{a}\rangle_{\star,1}\simeq -|\langle \hat{a}\rangle_{0}|^{2}\langle \hat{a}\rangle_{0}=-\alpha^{3}}\).
The simple approximation for the maximum of the absolute value of the amplitude is maximized for
\begin{align}
\label{eq:res-detu}
    \Delta  = \Delta_{\star,0}+\frac{4K^2\alpha^2}{\kappa^2}\left( 
    \color{red}n_\text{th}
    \color{black}
    +
    \color{blue}\frac{1}{2}
    \color{black}\right)(\Delta_{\star,0} - 2K\alpha^{2}-K/2) \, ,
\end{align}
where \(\Delta_{\star,0}=-K+K\alpha^{2}+2K(n_{\rm th}+\color{blue}1/2\color{black})\), and reads 
\begin{equation}
    |\langle \hat a\rangle_{\star} | \simeq \alpha\left(1-\frac{4K^2\alpha^2}{\kappa^2}\left( 
    \color{red}n_\text{th}
    \color{black}
    +
    \color{blue}\frac{1}{2}
    \color{black}\right)\right)\ .
    \label{eq:analytical_a}
\end{equation}
There are two corrections to the unperturbed amplitude $\alpha$: the terms in blue are due to the quantum commutation relations, and the terms in red are due to thermal effects.
We note that other authors have already derived more precise analytical equations for the amplitude (in absence of thermal noise)~\cite{drummond1980quantum,divincenzo2012nonlinear,meaney2014quantum}.

\subsection{Discussion}

The fact that an inclusion of noise leads to the same equations as the nonlinear damping model is confirmed by our fit of the latter model to data.
As covered in Sec.~\ref{sec:S_classical_solution}, the nonlinear damping is found to be $2\pi\times 5.02$ kHz, which is close to the value expected from this calculation.
Indeed, in absence of thermal noise $n_\text{th}=0$, Eq.~(\ref{eq:S_gamma_n}) gives $\gamma = 2K^2/\kappa = 2\pi\times 5.6$ kHz.

We can also use the approximate description of the resonance scenario embodied in Eqs.~(\ref{eq:res-detu})~and~(\ref{eq:analytical_a}) above to provide a value for the corresponding average photon number.
Given its equation of motion~(\ref{eq:motion-adag-a}), in the steady-state and on resonance, we have \(\langle \hat{a}^{\dagger}\hat{a}\rangle_{\star}=n_{\rm th}+\alpha (\langle \hat{a}\rangle_{\star} + \langle \hat{a^{\dagger}}\rangle_{\star})/2\).
Based on our previous perturbation series expansion \(\langle \hat{a}\rangle_{\star}\simeq \alpha[1-\gamma \alpha^{2}/\kappa]\), and hence
\begin{equation}
\begin{split}
  \langle \hat a^\dagger\hat a\rangle_{\star}&\simeq n_{\rm th} + \alpha^{2}\left(1-\frac{4K^2\alpha^2}{\kappa^2}\left(\color{red}n_\text{th}\color{black}+\color{blue}\frac{1}{2}\color{black}\right)\right)= \alpha^2\left(1+\frac{n_{\rm th}}{\alpha^2}- \frac{4K^2\alpha^2}{\kappa^2}\left(\color{red}n_\text{th}\color{black}+\color{blue}\frac{1}{2}\color{black}\right)\right)\ ,
\end{split}
\end{equation}
yielding, in the limit given by our assumptions (\ref{eq:S_approximation_small_K}) and (\ref{S_approximation_small_nth}), the following result
\begin{equation}
    \sqrt{\langle \hat a^\dagger\hat a\rangle_{\star}} \simeq \alpha\left(1+\frac{n_\text{th}}{2\alpha^2}-\frac{2K^2\alpha^2}{\kappa^2}\left(\color{red}n_\text{th}\color{black}+\color{blue}\frac{1}{2}\color{black}\right)\right)\ .
    \label{eq:analytical_n}
\end{equation}
The interplay of noise and Kerr nonlinearity leads to a reduction in $\sqrt{\langle \hat a^\dagger\hat a\rangle_{\star}}$ equal to half the reduction in $|\langle\hat a\rangle_{\star}|$ (Eq.~(\ref{eq:analytical_n})).
The steady-state of the driven-dissipative Kerr system differs thus from a coherent state, for which $\sqrt{\langle \hat a^\dagger\hat a\rangle_{\star}} = |\langle \hat a \rangle_{\star}|$, or a thermal coherent state, for which $\sqrt{\langle \hat a^\dagger\hat a\rangle_{\star}} \simeq |\langle \hat a \rangle_{\star}|+n_\text{th}/(2|\langle \hat a \rangle_{\star}|)$ if $n_\text{th}\ll |\langle \hat a \rangle_{\star}|$.
We can thus attribute half of the reduction of amplitude $|\langle \hat a\rangle_{\star}|$ to a reduction of photon number or energy of the oscillator. 

The subscript $\star$ is omitted in the rest of the manuscript where the resonant condition is implied through context.

\section{Wigner current}
\label{sec:S_wigner_current}
\subsection{Mathematical details}
\label{sec:S_wigner_current_math}
We follow Ref.~\cite{isar1996phase} to write the evolution of the Wigner function $W$ as a phase-space continuity equation
\begin{equation}
    \partial_t W + \vec\nabla \vec J = 0\ ,
\end{equation}
where $\nabla = \begin{pmatrix}
\partial_x\\
\partial_p
\end{pmatrix}$ and $\vec J = \begin{pmatrix}
J_x\\
J_p
\end{pmatrix}$ 
denotes the Wigner current.
This method has already been used to study the Kerr oscillator, however in the absence of a driving force~\cite{stobinska2008wigner,oliva2019quantum} or in a non-rotating frame~\cite{katz2008classical}.
The slightly different Hamiltonian of the Duffing oscillator, this time with both a driving force and damping has also been studied, however not in the rotating frame~\cite{friedman2017wigner}. 
Here we derive an expression for the Wigner current of the driven-dissipative Kerr oscillator in the rotating frame, characterized by the Hamiltonian
\begin{equation}
    \hat H/\hbar = \left(\Delta+\frac{K}{2}\right) \hat a^\dagger \hat a-\frac{K}{2}\left(\hat a^\dagger \hat a\right)^2 +i\epsilon(\hat a^\dagger- \hat a)\ .
\end{equation}
The slight difference in the form of nonlinearity with respect to Eq.~(\ref{eq:hamiltonian}) (allowed by the commutation relations) makes for a more favorable expression when Wigner-transforming the Hamiltonian to phase space coordinates $x,p$.
We introduce the phase-space operators as $\hat x,\hat p$ through
\begin{equation}
    \hat a = \frac{1}{\sqrt{2}}\left(\hat x+i\hat p\right)\ ,
\end{equation}
such that $[\hat x,\hat p]=i$, 
yielding
\begin{equation}
    \hat H/\hbar = \frac{1}{2}\left(\Delta+K\right) \left(\hat x^2+\hat p^2\right)-\frac{K}{8}\left(\hat x^2+\hat p^2\right)^2 +\epsilon\sqrt{2}\hat p\ ,
\end{equation}
omitting constant contributions.
To compute the Wigner current we first need the Wigner transform of the Hamiltonian, also known as the inverse of the Weyl transform, defined by~\cite{isar1996phase}
\begin{equation}
    H(x,p) = 2\int_{-\infty} ^{+\infty} \text d ze^{\frac{2ipz}{\hbar}}\left\langle x-z|\hat H|x+z\right\rangle\ ,
\end{equation}
which yields
\begin{equation}
    H(x,p)/\hbar = \frac{1}{2}\left(\Delta+K\right) \left(x^2+p^2\right)-\frac{K}{8}\left(x^2+p^2\right)^2 +\epsilon\sqrt{2}p\ ,
\end{equation}
omitting constant contributions.
It is easier to prove that $\hat H$ is the Weyl transform of $H(x,p)$ rather than the fact that $H(x,p)$
 is the Wigner transform of $\hat H$.
To do so, we may use the McCoy formula~\cite{mccoy}
\begin{equation}
    p^m x^n ~~ \longmapsto ~~ {1 \over 2^n} \sum_{r=0}^{n} {n \choose r} 
\hat x^r  \hat p^m  \hat x^{n-r}\ ,
\end{equation}
where $\longmapsto$ designates a Weyl transformation.
A consequence of this formula is that
\begin{equation}
\begin{split}
    x^n ~~ \longmapsto& ~~ \hat x^n\\
    p^n ~~ \longmapsto& ~~ \hat p^n\\
    2p^2x^2 ~~ \longmapsto& ~~ \frac12\left(\hat p^2\hat x^2+2 \hat x\hat p^2\hat x + \hat x^2 \hat p^2\right)\\
    &= \hat x^2\hat p^2 +\hat p^2\hat x^2 +1
\end{split}
\end{equation}
The two first relations prove the correspondence between the harmonic and driving terms.
Utilizing all three relations, we can demonstrate the correspondance between the Kerr terms in $\hat H$ and $H(x,p)$, up to a constant factor which plays no role in successive manipulations of $H(x,p)$.

The unitary evolution of the Wigner function, equivalent to the evolution of the state vector dictated by Schr\"odinger's equation, is given by~\cite{isar1996phase}
\begin{equation}
\label{eq:moyal}
\partial_t W(x,p,t)  =  \{\{H,W\}\}\  \equiv \frac{2}{\hbar} ~ H(x,p)\ \sin \left (
  {{\frac{1 }{2}}\left(\stackrel{\leftarrow }{\partial }_x \stackrel{\rightarrow
    }{\partial }_{p}-\stackrel{\leftarrow }{\partial }_{p}\stackrel{\rightarrow }{\partial
    }_{x}\right)} \right ) \ W(x,p,t),
\end{equation}
where $\{\{H,W\}\}$ is called the Moyal bracket.
The arrows above the partial derivatives indicates whether the term on the right or left should be differentiated.
For example:
\begin{equation}
    H(x,p)\stackrel{\leftarrow }{\partial }_xW(x,p,t) = \left(\partial_xH(x,p)\right)W(x,p,t)
\end{equation}
and 
\begin{equation}
    H(x,p)\stackrel{\rightarrow }{\partial }_xW(x,p,t) = H(x,p)\left(\partial_xW(x,p,t)\right)\ .
\end{equation}
Since the Hamiltonian only contains terms $x^np^m$ with $n+m\le 4$, we may write in this context
\begin{equation}
    \sin \left ({{\frac{1 }{2}}(\stackrel{\leftarrow }{\partial }_x \stackrel{\rightarrow}{\partial }_{p}-\stackrel{\leftarrow }{\partial }_{p}\stackrel{\rightarrow }{\partial}_{x})} \right ) = {\frac{1 }{2}}(\stackrel{\leftarrow }{\partial }_x \stackrel{\rightarrow}{\partial }_{p}-\stackrel{\leftarrow }{\partial }_{p}\stackrel{\rightarrow }{\partial}_{x})- {\frac{1 }{3!}\frac{1 }{2^3}}(\stackrel{\leftarrow }{\partial }_x \stackrel{\rightarrow}{\partial }_{p}-\stackrel{\leftarrow }{\partial }_{p}\stackrel{\rightarrow }{\partial}_{x})^3\ .
\end{equation}

The Moyal bracket for the harmonic part of the Hamiltonian writes
\begin{equation}
\begin{split}
&2\ \frac{1}{2} \left(\Delta+K\right)\left(x^2+p^2\right){\frac{1 }{2}}(\stackrel{\leftarrow }{\partial }_x \stackrel{\rightarrow}{\partial }_{p}-\stackrel{\leftarrow }{\partial }_{p}\stackrel{\rightarrow }{\partial}_{x})W(x,p,t)\\
=&\left(\Delta+K\right) (x \stackrel{\rightarrow}{\partial }_{p}-p\stackrel{\rightarrow }{\partial}_{x})W(x,p,t)\\
=&-\begin{pmatrix}
\partial_x\\
\partial_p
\end{pmatrix}
\underbrace{
\begin{pmatrix}
p\\-x
\end{pmatrix}\left(\Delta+K\right)W(x,p,t)}_{\vec J_\text{harmonic}}\ .
\end{split}
\label{eq:S_J_harmonic}
\end{equation}
For the drive term
\begin{align}
&2\epsilon\sqrt{2}p{\frac{1 }{2}}(\stackrel{\leftarrow }{\partial }_x \stackrel{\rightarrow}{\partial }_{p}-\stackrel{\leftarrow }{\partial }_{p}\stackrel{\rightarrow }{\partial}_{x})W(x,p,t)\\
=&-\epsilon\sqrt{2}\stackrel{\rightarrow }{\partial
    }_{x}W(x,p,t)\\
=&-\begin{pmatrix}
\partial_x\\
\partial_p
\end{pmatrix}
\underbrace{\begin{pmatrix}
1\\ 0
\end{pmatrix}\epsilon\sqrt{2}W(x,p,t)}_{\vec J_\text{drive}}\ .
\end{align}
\begin{figure}[htb]
  \begin{centering}
    \includegraphics[width=0.9\textwidth]{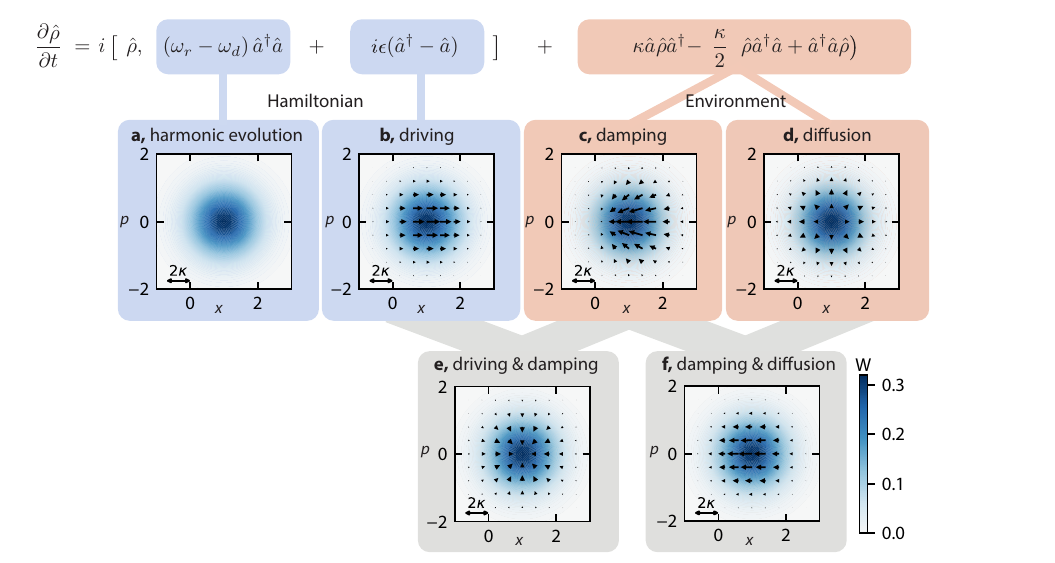}
  \par\end{centering}
\caption{
\textbf{Wigner current in a resonantly-driven harmonic oscillator.}
We overlay the Wigner function of the steady-state (a coherent state), with the Wigner current associated with the different terms of the Lindblad equation.
The sum of these vector fields $\vec J$  plays the role of a current in a continuity equation $\partial W/\partial t = \vec\nabla \vec J$ equivalent to the Lindblad equation.
\textbf{a}, On resonance $\omega_\text{r}=\omega_\text{d}$, there is no harmonic evolution in the rotating frame.
\textbf{b}, The driving moves the Wigner function towards increasing $x$.
\textbf{c}, The damping moves the Wigner function towards the origin.
\textbf{d}, The diffusion term (quantum noise) expands the Wigner function in all directions.
\textbf{e}, Since the drive acts along the $x-$axis and the damping acts more radially, these two terms do not counteract each other.
Rather, they move the Wigner function towards a point in phase space, the classical solution to the problem.
This is counteracted by the diffusion (quantum noise).
Together, these effects give the coherent state its finite size in phase space.
\textbf{f}, Another way of showing this is to add the effect of the damping and diffusion, which yields a current acting in the $x$-axis rather than radially, which counteracts the driving.
}
\label{fig:S_wigner_current_resonant_HO}
\end{figure}
\begin{figure}[htb]  
  \begin{centering}
    \includegraphics[width=0.9\textwidth]{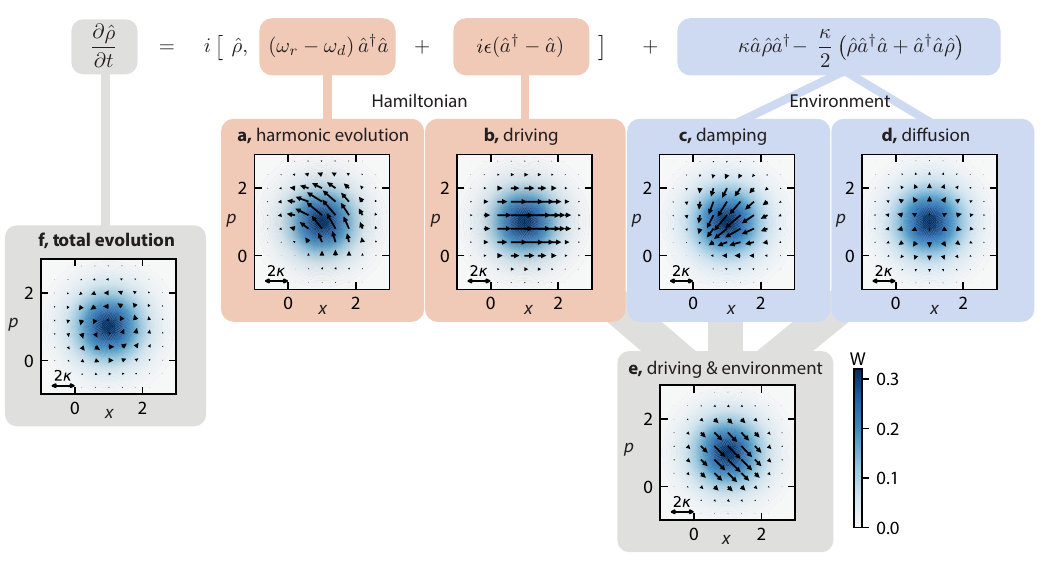}
  \par\end{centering}
  \caption{
\textbf{Wigner current in an off-resonantly driven harmonic oscillator.}
We overlay the harmonic \textbf{a}, drive \textbf{b}, damping \textbf{c}, and diffusion (quantum noise) \textbf{d} Wigner currents on the Wigner distribution of the steady-state.
\textbf{e}, Contrary to the resonant case, the sum of diffusion and damping no longer cancel the drive.
Instead, these currents point away from the direction of the harmonic evolution, such that the net total current is 0.
\textbf{f}, The remaining current moves points of the Wigner distribution along lines of equal probability such that $\partial W/\partial t=0$\ .
}
\label{fig:S_wigner_current_OFF_resonant_HO}
\end{figure}
The first order derivatives of the Moyal bracket applied to the nonlinearity write
\begin{equation}
\begin{split}
&-2\frac{K}{8}\left(x^2+p^2\right)^2{\frac{1 }{2}}(\stackrel{\leftarrow }{\partial }_x \stackrel{\rightarrow}{\partial }_{p}-\stackrel{\leftarrow }{\partial }_{p}\stackrel{\rightarrow }{\partial}_{x})W(x,p,t)\\
=&-\frac{K}{2}\left(x^2+p^2\right)(x \stackrel{\rightarrow
    }{\partial }_{p}-p\stackrel{\rightarrow }{\partial
    }_{x})W(x,p,t)\\
=&-\begin{pmatrix}
\partial_x\\
\partial_p
\end{pmatrix}
\underbrace{\begin{pmatrix}
p\\-x
\end{pmatrix}\left(-\frac{K}{2}\left(x^2+p^2\right)\right)W(x,p,t)}_{\vec J_\text{Kerr,1}}\ .
\end{split}
\label{eq:S_J_Kerr1}
\end{equation}
And the higher order derivatives of the Moyal bracket applied to the nonlinearity yield
\begin{equation}
\begin{split}
&-2\frac{K}{2}\frac{1}{4}\left(x^2+p^2\right)^2\left(- {\frac{1 }{3!}\frac{1 }{2^3}}(\stackrel{\leftarrow }{\partial }_x \stackrel{\rightarrow}{\partial }_{p}-\stackrel{\leftarrow }{\partial }_{p}\stackrel{\rightarrow }{\partial}_{x})^3\right)W(x,p,t)\\
=&-\genfrac{(}{)}{0pt}{}{{\partial}_{x}}{{\partial }_{p}}\underbrace{\frac{K}{24}\genfrac{(}{)}{0pt}{}{p(3{\partial }_{xx}+{\partial }_{pp})-x({\partial }_{xp}+{\partial }_{px})}{-x({\partial }_{xx}+3{\partial }_{pp})+p({\partial }_{xp}+{\partial }_{px})}W(x,p,t)}_{\vec J_\text{Kerr,2}}\ ,
\end{split}
\label{eq:S_J_Kerr2}
\end{equation}
which in our regime of parameters has a negligibly small contribution to the total current.
The non-unitary evolution of the Wigner function, equivalent to the Lindblad equation of Eq.~(\ref{eq:lindblad}), is given by~\cite{friedman2017wigner}
\begin{equation}
    \partial_tW = \{\{H,W\}\}\ -
\begin{pmatrix}
\partial_x\\
\partial_p
\end{pmatrix}
\left(\underbrace{-\frac{\kappa}{2}
\begin{pmatrix}
x\\p
\end{pmatrix}W}_{\vec J_\text{damping}}
\underbrace{-\frac{\kappa}{2}\left(n_\text{th}+\frac{1}{2}\right)
\begin{pmatrix}
\partial_x\\\partial_p
\end{pmatrix}W}_{\vec J_\text{diffusion}}
\right)\ .
\end{equation}

\begin{figure}[htb]
\begin{centering}
\includegraphics[width=0.9\textwidth]{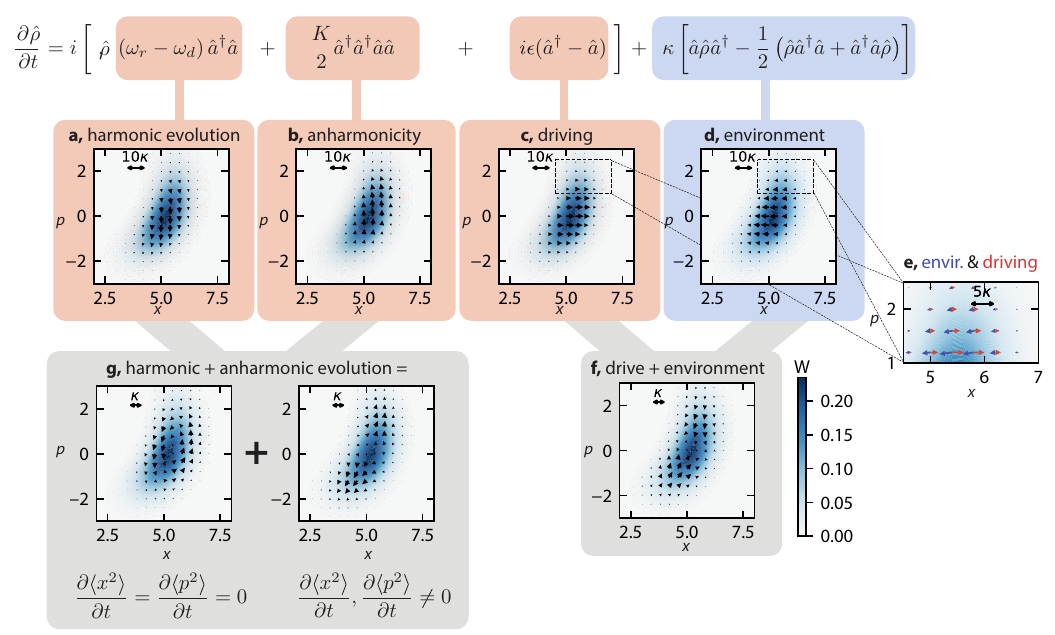}
\par\end{centering}
\caption{
\textbf{Wigner current in the steady-state of the driven Kerr oscillator.}
We overlay the contributions of the harmonic \textbf{a}, anharmonic \textbf{b}, drive \textbf{c}, and environmental \textbf{d} terms of the Lindblad equation on the Wigner function of Fig.~4\textbf{e}.
\textbf{e}, Zoom-in showing both environmental and driving contributions.
Whilst the drive acts on the $x-$axis, the environment acts more radially. 
They only partially compensate each other, and the result of summing these currents is shown in \textbf{f}.
Together, they tend to diminish the spread in phase of the Wigner function.
\textbf{g}, This is compensated by the anharmonicity which tends to increase the spread in phase.
To make this clear, we have divided the sum of the harmonic and anharmonic currents into two contributions.
In \textbf{g}-left we have shown the part of the current which preserves the spread in position $\partial\langle x^2\rangle/\partial t=\partial \langle p^2\rangle/\partial t=0$.
This corresponds to the total Wigner current.
In \textbf{g}-right we have shown the part of the current which increases the spread in position, and exactly compensates the combined effect of the drive and environment.
The latter is plotted in Fig.~4\textbf{c}.
}
\label{fig:S_wigner_current_Kerr}
\end{figure}

These expressions are utilized in the discussion surrounding Fig.~4.
We supplement Fig.~4 by the discussion below, where we provide details on the construction of the figure and further arguments in favor of its interpretation.
We also provide in Fig.~\ref{fig:S_wigner_current_Kerr} a detailed plot of the Wigner currents for the steady-state shown in Fig.~4\textbf{c}.
Finally, we plot in Fig.~\ref{fig:S_wigner_current_resonant_HO} and Fig.~\ref{fig:S_wigner_current_OFF_resonant_HO} the Wigner currents for the pedagogical cases of a resonantly and off-resonantly driven harmonic oscillators respectively.
\subsection{Supplementary information for Fig.~4}

In Fig.~4\textbf{b}, we show a coherent state, of amplitude $\alpha$, and the state which results from evolving that coherent state under the Kerr effect $\hat H/\hbar =  \lambda K\hat a^\dagger\hat a - (K/2)\hat a^\dagger\hat a^\dagger\hat a\hat a$ for a time $t=1/(45K)$.
This unitary evolution is computed using QuTiP~\cite{johansson2012qutip,JOHANSSON20131234}.
To ensure that the state is centered around the $x$ axis in phase space, we add a harmonic evolution characterized by the coefficient $\lambda = 15$.
After evolution of the coherent state due to the Kerr effect, the average photon number remains unchanged.
This is theoretically expected since the Hamiltonian dictating this evolution $\hat H = \hbar \lambda K\hat a^\dagger\hat a - \hbar (K/2)\hat a^\dagger\hat a^\dagger\hat a\hat a$ commutes with the photon number operator $\hat a^\dagger \hat a$.

The amplitude $|\langle\hat a \rangle|$, or the distance between the center of mass of the distribution and the origin in phase space, decreases.
In Fig.~4\textbf{b}, the decrease in amplitude is of 0.34 percent, which translates to a considerable 0.33 dB difference in $|S_\text{21}|$.
We have called this \textit{effect A}.

\begin{figure}[htb]
  \begin{centering}
    \includegraphics[width=0.9\textwidth]{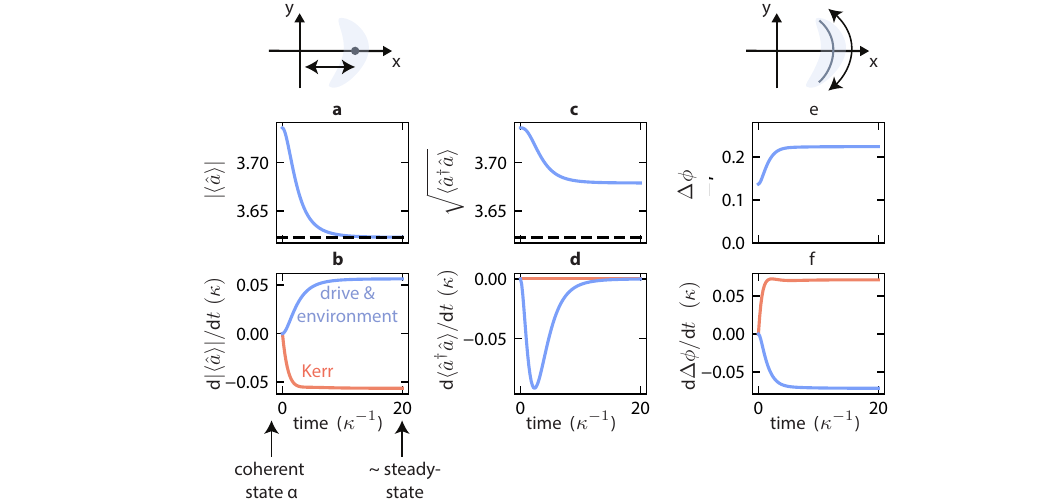}
  \par\end{centering}  
\caption{
\textbf{Evolution of the driven-dissipative Kerr oscillator starting in a coherent state.}
We compute the time-evolution of the density matrix following Eq.~(\ref{eq:lindblad}) in QuTiP~\cite{johansson2012qutip,JOHANSSON20131234}.
The starting state is taken to be a coherent state of amplitude $\alpha$ (see Fig.~4\textbf{a}), the amplitude corresponding to the classical solution to the problem.
The simulation is carried out with a power $P_\text{in} = -124$, until the steady-state shown in Fig.~4\textbf{c} is reached.
The time evolution of various observables are plotted over time in order to further illustrate the amplitude reduction mechanism shown in Fig.~4.
We see in \textbf{b} that the Kerr causes $|\langle \hat a \rangle|$ to reduce over time (negative time-derivative) whereas the drive and environment counteracts this effect (positive time-derivative) until equilibrium is reached at the steady-state (the total time-derivative is 0).
This causes the reduction of $|\langle \hat a \rangle|$ shown in \textbf{a}.
The dashed line corresponds to the steady-state value for $|\langle \hat a \rangle|$ in \textbf{a} and \textbf{c}.
In \textbf{c}, we plot the square of the average photon number, and see that a reduction in photon number accounts for half of the total reduction in $|\langle \hat a \rangle|$.
From \textbf{d}, we find that the photon number is conserved under the Kerr evolution (derivative is 0), whereas the interaction of the drive and environment is at the origin of the decrease.
The phase-space interpretation of this effect is that through an increase in the phase variance, the environmental Wigner current is no longer parallel to the driving current (see Fig.~4\textbf{c}), creating a net current of the probability towards the origin (\textit{i.e.} a reduction in photon number).
Since the amplitude of the damping current is reduced closer the origin (\textit{i.e.} is smaller for lower photon numbers), whereas the drive is not, equilibrium is found when the probability gets closer to the origin.
In \textbf{e}, we plot the evolution of the variance in phase $\Delta \phi$.
Its increase is shown to be due to the Kerr effect in \textbf{f}, and countered by the interaction of drive and environment.
}
\label{fig:S_coherent_to_steadystate_evolution}
\end{figure}

As a result of the deformation of the Wigner distribution, the drive is less affective at countering the environmental effects (noise and damping).
If we call $\vec J_\text{env} = \vec J_\text{damping}+\vec J_\text{diffusion}$, and integrate the absolute value of this current, and the driving current over one of the distribution of Fig.~4\textbf{b}, we find $\iint |\vec J_\text{env}|\text dx\text dp = \iint |\vec J_\text{drive}|\text dx\text dp$.
Whilst this is true for the coherent state and the state deformed by the Kerr effect, we obtain a different result if we project the driving current in the same direction as the environmental effects.
For a coherent state, we have $\iint |\vec J_\text{env}|\text dx\text dp = \iint |\vec J_\text{drive}\cdot \frac{\vec J_\text{env}}{|\vec J_\text{env}|}|\text dx\text dp$.
So the drive is aligned with the environmental effects and exactly counters them.
However, for the deformed state $\iint |\vec J_\text{env}|\text dx\text dp = 6.10$ MHz, and $\iint |\vec J_\text{drive}\cdot \frac{\vec J_\text{env}}{|\vec J_\text{env}|}|\text dx\text dp = 6.07$ MHz.
So after effect of the Kerr, whilst the drive matches the environmental effects in absolute magnitude, in the direction of the environmental current the driving current is smaller.
The Wigner distribution thus tends to move towards the origin in phase space, hence reducing the total photon number.
We have called this \textit{effect B}.
This is further illustrated if we track the evolution of various observables as a coherent state of amplitude $\alpha$ evolves to the steady-state following Eq.~(\ref{eq:lindblad}), see Fig.~\ref{fig:S_coherent_to_steadystate_evolution}.
We look at the amplitude $|\langle\hat a\rangle| = |\text{Tr}(\hat a\hat \rho)|$, the square root of the number of photons $\sqrt{\langle\hat a^\dagger\hat a\rangle} = \sqrt{\text{Tr}(\hat a^\dagger \hat a\ \hat \rho)}$ and the variance of the phase $\Delta\varphi = \sqrt{\text{Tr}(\hat \varphi^2\hat \rho)-\text{Tr}(\hat \varphi\hat \rho)^2}$.
Additionally, we look at the time derivative of these observables induced by various terms of Eq.~(\ref{eq:lindblad}).

We distinguish the influence of different terms by computing the change in density matrix induced by drive and damping 
\begin{equation}
\begin{split}
    \left[\frac{\partial\hat\rho}{\partial t}\right]_{\kappa,\epsilon} =& -i\left[i\epsilon(\hat a^\dagger-\hat a),\hat \rho\right]\\
     &+ \kappa (n_\text{th}+1) D(\hat a )\hat \rho+ \kappa n_\text{th} D(\hat a^\dagger )\hat \rho\ ,
\end{split}
\end{equation}
and by the harmonic and Kerr evolution
\begin{equation}
    \left[\frac{\partial\hat\rho}{\partial t}\right]_{K} = -i\left[\Delta\hat a^\dagger \hat a -\frac{K}{2}\hat a^\dagger\hat a^\dagger \hat a \hat a,\hat \rho\right]\ .
\end{equation}
Given a change in density matrix $\tfrac{\partial\hat\rho}{\partial t}$, we can compute the change in amplitude from
\begin{equation}
\begin{split}
    \frac{\partial|\langle\hat a\rangle|}{\partial t}&=\frac{\partial}{\partial t}\sqrt{\langle\hat a\rangle^*\langle\hat a\rangle}\\
    &=\frac{1}{2}\left(\left(\frac{\partial}{\partial t}\langle\hat a\rangle\right)^*\langle\hat a\rangle+\langle\hat a\rangle^*\left(\frac{\partial}{\partial t}\langle\hat a\rangle\right)\right)/|\langle\hat a\rangle|\ ,
\end{split}
\end{equation}
where 
\begin{equation}
\begin{split}
    \frac{\partial}{\partial t}\langle\hat a\rangle &= \frac{\partial}{\partial t}\text{Tr}(\hat a \hat \rho)\\
    &= \text{Tr}(\hat a \frac{\partial\hat\rho}{\partial t})\ .
\end{split}
\end{equation}
For the photon number
\begin{equation}
    \frac{\partial}{\partial t}\langle\hat a^\dagger\hat a\rangle= \text{Tr}(\hat a^\dagger\hat a \frac{\partial\hat\rho}{\partial t})\ .
\end{equation}
And for the variance in the phase
\begin{equation}
    \frac{\partial}{\partial t}\Delta\varphi = \frac{\text{Tr}(\hat \varphi^2\frac{\partial\hat\rho}{\partial t})-2\text{Tr}(\hat \varphi\hat \rho)\text{Tr}(\hat \varphi\frac{\partial\hat\rho}{\partial t})}{2\Delta\varphi}\ .
\end{equation}
\begin{figure}[htb]
  \begin{centering}
    \includegraphics[width=0.9\textwidth]{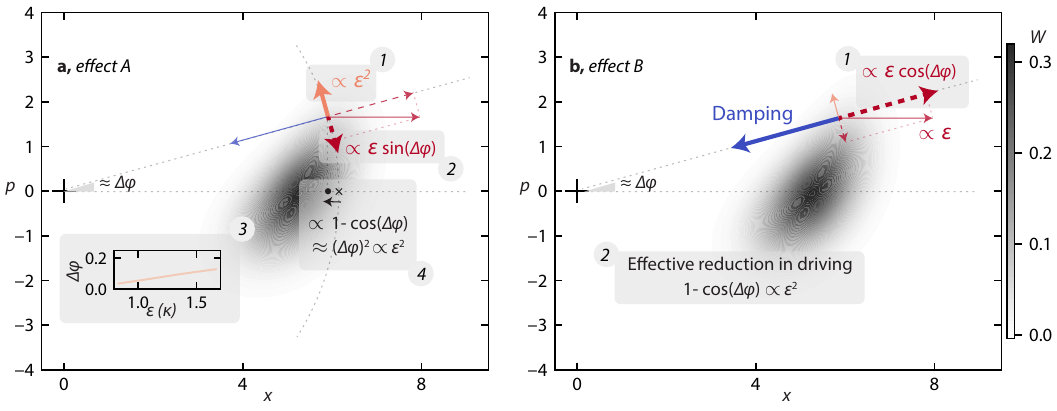}
  \par\end{centering}
\caption{
    \textbf{Origin of the nonlinear nature of the damping.}
    Following a unitary evolution of a coherent state due to the Kerr effect, the state acquires an uncertainty in phase characterized by $\Delta \varphi$.
    We show in both panels the Wigner function of Fig.~4\textbf{b} illustrating this effect.
    The Wigner currents are drawn with arbitrary lengths for pedagogical purposes and do not reflect their true values.
    \textbf{(a}
    -1\textbf{)} A point in phase space which has acquired a phase offset $\Delta \varphi$ will be subject to the Kerr and harmonic evolution scaling with the amplitude squared, or in other words $\epsilon^2$.
    (2) When the damping and driving are active, the steady-state is reached when the Kerr current is compensated by the tangential component of the drive which scales with $\epsilon \sin (\Delta \varphi)$.
    (3) For an equilibrium to occur, and assuming the dephasing is small $\sin (\Delta \varphi) \simeq \Delta \varphi$, the dephasing should scale linearly with $\epsilon$.
    In the inset we show a simulation of the phase uncertainty $\Delta \varphi = \Delta \phi-1/n$ which is acquired with increasing driving.
    To compute $\Delta \varphi$, we have subtracted the phase uncertainty of a coherent state $1/n = 1/\langle\hat a^\dagger\hat a\rangle$, which does not contribute to the amplitude damping, to the total uncertainty in phase $\Delta \phi$. 
    As expected, $\Delta \varphi$ increases linearly with $\epsilon$.
    (4) By projecting the equilibrium point on the axis of the center of mass of the distribution (COM), we find a contribution to the reduction of the COM proportional to $\epsilon^2$ . This scaling is consistent with our experimental and theoretical findings, and helps to understand why \textit{effect A} leads to an apparent damping which is nonlinear in $\epsilon$.
    \textbf{b}
    When we look at the balance of currents in the radial direction, we find that the phase offset $\Delta \varphi$ causes the contribution of the driving current to be reduced by a factor $(\epsilon - \epsilon \cos (\Delta \varphi) )/\epsilon \simeq \epsilon^2$.
    This is consistent with the reduction in photon number proportional to the amplitude squared, and helps to understand why \textit{effect B} leads to an apparent nonlinear damping.
    }
\label{fig:S_nonlinear_intuition}
\end{figure}
\newpage
Consistently with the explanations surrounding Fig.4, we observe that the Kerr effect causes a decrease of the amplitude (Fig.~\ref{fig:S_coherent_to_steadystate_evolution}\textbf{b}) \textit{(effect A)}.
And this effect is eventually compensated by the combined effect of the drive and environment.
We also observe that the average number of photons decreases due to the interaction of drive and damping (Fig.~\ref{fig:S_coherent_to_steadystate_evolution}\textbf{f}), and this effect seems to follow the increase in phase variance (Fig.~\ref{fig:S_coherent_to_steadystate_evolution}\textbf{f}) \textit{(effect B)}.

Whilst the explanations above and surrounding Fig.~4 help in understanding the physical mechanism behind the dephasing which manifests in the same way as damping, they do not elucidate its nonlinear nature.
To understand the latter, we turn to estimating the equilibrium of Wigner currents (see Fig.~\ref{fig:S_nonlinear_intuition} for a visual point of reference).
We consider the state of Fig.~4\textbf{b}, after the Kerr effect has acted on the coherent state.
Assuming we are driving at the new resonance frequency\footnote[1]{We note that, in the absence of thermal fluctuations, \(n_{\rm th}< 0.05\), as we expect in our experiment, the mismatch between the resonance detuning \(\Delta = K\alpha^{2}\) as contemplated by a classical description of the damped driven Kerr oscillator and its approximated counterpart value of Eq.~(\ref{eq:res-detu}) that derives from our quantum description, is lesser than a \(4\%\) for \(P_{\rm in} \leq -124\) dBm.
  Thus, using Eq.~(\ref{eq:res-detu}) instead, entails almost no change in the steady-state of the Wigner currents with its overall features remaining the same.} $\Delta = K\alpha^2$, we determine the scaling of currents for a point which has undergone a characteristic amount of dephasing $\Delta\varphi$.
We consider a point in phase space which has a distance from the origin given by $\langle \hat x\rangle + \Delta x$, where $\Delta x = 1/\sqrt{2}$ is the uncertainty in position of a coherent state and $\langle \hat x\rangle = \epsilon/(\sqrt{2}\kappa)$.
We find that the Kerr and harmonic evolution Wigner current $\vec J_\text{Kerr,1}+\vec J_\text{harmonic}$ scales with $\epsilon^2$, and points in a perpendicular direction to the damping.
Our numerics show that in practice the second contribution to the Kerr current $\vec J_\text{Kerr,2}$ is negligibly small, and we see in Fig.~4\textbf{c} that the influence of quantum noise at large dephasing is also negligible.
In this discussion we will simplify the steady-state condition as requiring the orthogonal Wigner currents of Kerr/harmonic evolution and damping to be matched by the drive which is oriented horizontally.
As illustrated in Fig.~\ref{fig:S_nonlinear_intuition} this condition leads to an understanding of the nonlinear evolution of the apparent damping, which decreases the total amplitude $\alpha$ by a factor $\epsilon ^2$, resulting in the nonlinear scaling of amplitude $\alpha\epsilon ^2\propto \alpha^3$ found both experimentally and theoretically.

\section{Quadrature squeezing}

\begin{figure}[htb]
  \begin{centering}
    \includegraphics[width=0.9\textwidth]{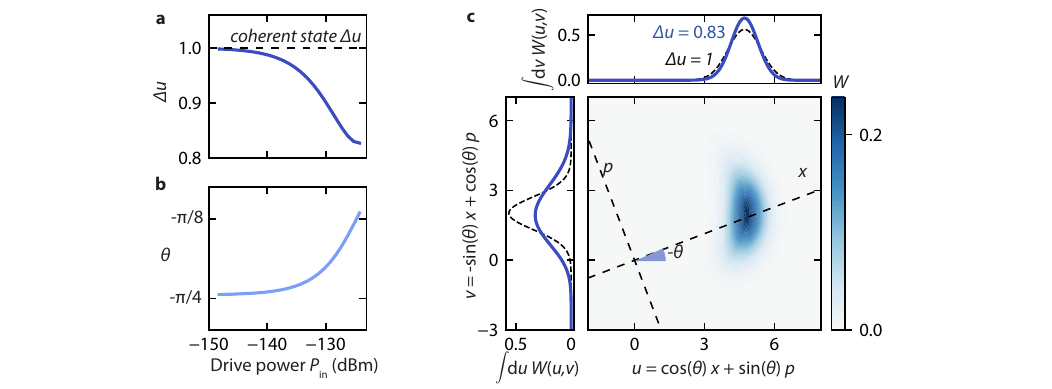}
  \par\end{centering}
\caption{\textbf{Quadrature squeezing.}
The angle $\theta$ defines the orthonormal coordinate set $u,v$ with $u = \cos(\theta)x+\sin(\theta)p$ where the uncertainty in $\hat u$ named $\Delta u$ is minimized.
For each driving power, we plot the minimum $\Delta u$ in \textbf{a} and the angle which achieves this minimum in \textbf{b}.
The uncertainty $\Delta u$ is expressed relative to the quadrature uncertainty of a coherent state shown as a dashed line.
\textbf{c}, Wigner function plotted in the $u,v$ basis for $P_\text{in} =-124.2$ dBm where the smallest $\Delta u$ is achieved.
On the left and on top, we plot the marginal distributions for $u$ and $v$ respectively (blue), overlayed with the distributions of a coherent state (dashed lines) with the same amplitude $\langle \hat a\rangle$ as the plotted Wigner function.
In this case, the uncertainty $\Delta u$ for the steady-state is 83 percent of the uncertainty one would obtain for a coherent state, demonstrating quadrature squeezing.
}
\label{fig:S_squeezing}
\end{figure}

We determine the degree of squeezing of the steady-state solutions to Eq.~(\ref{eq:lindblad}) for different powers.
We consider squeezing along the orthonormal coordinate set $u,v$ with $u = \cos(\theta)x+\sin(\theta)p$.
The coordinate $u$ corresponds to the operator $\hat u = e^{i\theta}\hat a^\dagger + e^{-i\theta}\hat a$ for which we compute the uncertainty $\Delta u$ for varying $\theta$.
As shown in Fig.~\ref{fig:S_squeezing}, for each measured power, there is a angle $\theta$ for which the uncertainty is below the uncertainty of a coherent state, demonstrating quadrature squeezing.

\section{Device fabrication and setup}
\label{sec:S_fab_setup}
      
\subsection{Fabrication}
The device shown in Fig.~1 is fabricated in two steps~\cite{yanai2019observation}. 
First, we fabricate the input/output waveguide structures, meandering inductor and capacitor.
On a chip of high-resistivity silicon, cleaned in solutions of RCA-1, Piranha, and buffered hydrofluoric acid (BHF), we sputter 60 nm of molybdenum-rhenium (MoRe).
A three layer mask (S1813/W(tungsten)/PMMA-950) is then patterned using electron-beam lithography, and is used in etching the MoRe by SF6/He plasma.
The mask is finally stripped using PRS 3000.

Secondly, we fabricate the Josephson junctions using the Dolan bridge technique~\cite{dolan1977offset}.
We first pattern a Methyl-methacrylate (MMA) /Polymethyl-methacrylate (PMMA) resist stack with e-beam lithography.
After development of the resist, and to ensure a good contact between the aluminum of the junctions and the MoRe, we clean the sample with an oxygen plasma and BHF. 
Evaporation of two aluminum layers (30 nm and then 50 nm thick) under two angles ($\pm$ 11 degrees), interposed by an oxidization of the first aluminum layer, forms the junctions.
Removal of the resist mask in N-Methyl-2-pyrrolidone (NMP) at 80 degrees Celsius completes the sample fabrication.

\subsection{Experimental setup}
\label{sec:experimental-setup}

\begin{figure}[htb]
  \begin{centering}
    \includegraphics[width=0.96\textwidth]{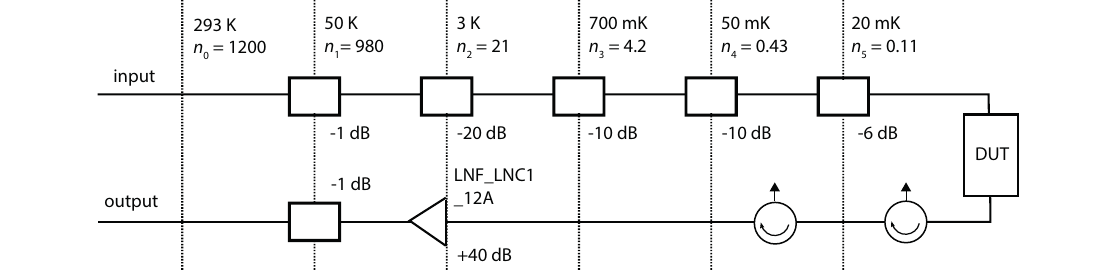}
   \par\end{centering}
\caption{
	\textbf{Cryogenic wiring diagram.}
	On the input side, at each temperature stage (corresponding to a plate of our dilution refrigerator) the signal passes through a thermalized attenuator.
	This reduces the noise photon occupation $n_i$.
	On the output side, the signal passes through two circulators which isolate the device from noise from the HEMT amplifier at 3 K.
	}
	\label{fig:S_8}
      \end{figure}

The edges of the chip are wire-bonded to a printed circuit board (PCB) and the chip is placed in a copper box thermally anchored to the 20 milliKelvin stage of a dilution refrigerator.
The input to the device is wired through the PCB to the output of a room-temperature vector network analyzer (VNA), the signal coming from the VNA is attenuated at each plate of the dilution refrigerator.
The cryogenic wiring is detailed in Fig.~\ref{fig:S_8}.

The noise photon occupation $n_i$ of a coaxial cable microwave mode, at a plate $i$, with frequency $\omega$ is given by~
\begin{equation}
\begin{split}
	n_i(\omega) = \frac{n_{i-1}(\omega)}{A_i}+\frac{A_i-1}{A_i}n_\text{BE}(T_i,\omega),\\
	n_\text{BE}(T,\omega) = \frac{1}{e^{\frac{\hbar\omega}{k_\text{B} T}-1}}
\end{split}
\end{equation}
where $n_{i-1}(\omega)$ is the occupation for the plate with the next higher temperature, $n_0=n_\text{BE}(293 \, {\rm Kelvin},\omega)$, $A_i$ is the attenuation 
at the plate $i$ with temperature $T_i$, such that for an attenuator with value -10 dB, $A = 10$.
%




The noise temperature of modes coming towards the device from the input is then $n_\text{th,in}<0.108$, which is only an upper bound as additional attenuation in the cabling and connectors has not been taken into account.
For modes coming towards the device from the output wiring, the noise is given by the 50 $\Omega$ Johnson noise in the isolator $n_\text{th,out}\simeq0$. 
The thermal occupancy of the resonator $n_\text{th}$ is the sum of the occupancy of each thermal bath to which the resonator is coupled to, weighted by the coupling to each bath.
Since the device is over-coupled to the feedline, the average occupancies of modes coming towards the device from both directions is a good approximation of the resonator occupancy: $n_\text{th}<0.05\ll1/2$.
The output of the device is wired to the input of the VNA after being amplified with a high-electron-mobility transistor (HEMT) amplifier at $\sim$4 Kelvin, and a room-temperature amplifier.

\subsection{Frequency-dependence of damping}
\begin{figure*}[t!]
\includegraphics[width=0.48\textwidth]{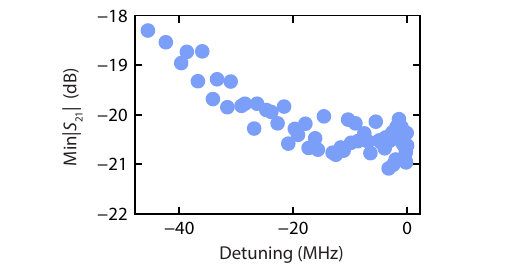}
\caption{
	\textbf{Minimum of transmission as a function of detuning. }
	The detuning is defined as the difference between the circuit's resonance frequency -- modified by the magnetic flux traversing the SQUID loop -- and the resonance frequency used in the rest of the data presented in this work.
	}
	\label{fig:flux_sweep}
	\end{figure*}

By sweeping the current in a coil located under the device we can change the static magnetic field traversing the device.
Through this field, we are able to tune the Josephson inductance of the SQUID and hence the resonance frequency of the circuit.
In the data presented in this paper, we operate at the first order magnetic field insensitive point of the SQUID - also known as flux "sweet-spot".
In Fig.~\ref{fig:flux_sweep}, we show our analysis of the data presented in Ref.~\cite{yanai2019observation}, where $S_{21}$ is recorded as the flux through the SQUID is changed.
This data set is acquired at a power of -138 dBm, far below the power at which the apparent nonlinear damping manifests.
We find a change in the depth of the resonance peak as a function of detuning from the sweet-spot.
However, the detuning required to match the change induced by power far exceeds any change in resonance frequency occurring through the Kerr nonlinearity.
Indeed, the shift induced by driving the Kerr oscillator remains under 2 MHz (see Fig.~\ref{fig:S_fitted_min_fmin}) but the minimum changes by 4 dB, whereas the changes in Min$|S_{21}|$ shown in Fig.~\ref{fig:flux_sweep} in the detuning range of a few MHz remains smaller than 1 dB, dominated by experimental noise.
This means for example that any frequency-dependence in the measurement port admittances as seen from the resonator is not responsible for the non-linear damping discussed in this work.

\FloatBarrier
\normalem
\bibliography{library}

\end{document}